\documentclass{aa}
\usepackage{graphicx}
\usepackage{txfonts}
\def\kms {\hbox{${\rm km\,s}^{-1}$}}

\def\Jypb {\hbox{Jy\,beam$^{-1}$}}

\def\as {\hbox{$^{\prime\prime}$}}

\def\ad {\hbox{$^{\circ}$}}

\def\ffas {\hbox{$\,.\!\!^{\prime\prime}$}}

\def\ffad {\hbox{$\,.\!\!^{\circ}$}}

\def \la{\mathrel{\mathchoice   {\vcenter{\offinterlineskip\halign{\hfil
$\displaystyle##$\hfil\cr<\cr\sim\cr}}}
{\vcenter{\offinterlineskip\halign{\hfil$\textstyle##$\hfil\cr
<\cr\sim\cr}}}
{\vcenter{\offinterlineskip\halign{\hfil$\scriptstyle##$\hfil\cr
<\cr\sim\cr}}}
{\vcenter{\offinterlineskip\halign{\hfil$\scriptscriptstyle##$\hfil\cr
<\cr\sim\cr}}}}}
\def \ga{\mathrel{\mathchoice   {\vcenter{\offinterlineskip\halign{\hfil
$\displaystyle##$\hfil\cr>\cr\sim\cr}}}
{\vcenter{\offinterlineskip\halign{\hfil$\textstyle##$\hfil\cr
>\cr\sim\cr}}}
{\vcenter{\offinterlineskip\halign{\hfil$\scriptstyle##$\hfil\cr
>\cr\sim\cr}}}
{\vcenter{\offinterlineskip\halign{\hfil$\scriptscriptstyle##$\hfil\cr
>\cr\sim\cr}}}}}

\begin{document}

   %\title{A collimated, pulsed, and precessing jet associated with an O-type protostar in G26.50+0.28}
   \title{ALMA and VLBA views on the outflow associated with an O-type protostar in G26.50+0.28}

   \titlerunning{ALMA and VLBA views on the outflow in G26.50+0.28}

   \author{Gang Wu
          \inst{1}\fnmsep\inst{2}
          \and
          Christian Henkel
          \inst{1}\fnmsep\inst{3}\fnmsep\inst{2}
          \and
          Ye Xu
          \inst{4}
          \and
          Andreas Brunthaler
          \inst{1}
          \and
          Karl M. Menten
          \inst{1}
          \and
          Keping Qiu
          \inst{5}
          \and
          Jingjing Li
          \inst{4}
          \and
          Bo Zhang
          \inst{6}
          \and
          Jarken Esimbek
          \inst{2}
           }

   \institute{Max-Planck-Institut f\"{u}r Radioastronomie, Auf dem H\"{u}gel 69, 53121, Bonn, Germany\\
   \email{gwu@mpifr-bonn.mpg.de}
   \and Xinjiang Astronomical Observatory, CAS  150, Science 1-Street Urumqi, Xinjiang 830011, China
   \and Astronomy Department, King Abdulaziz University, PO Box 80203, Jeddah 21589, Saudi Arabia
   \and Purple Mountain Observatory, CAS, No.8 Yuanhua Road, Qixia District, Nanjing 210034, China
   \and School of Astronomy and Space Science, Nanjing University, 163 Xianlin Avenue, Nanjing 210023, China
   \and Shanghai Astronomical Observatory, Chinese Academy of Sciences, 80 Nandan Road, Shanghai 200030, China
   }

   \date{Received ...; accepted ...}

  \abstract
   {Protostellar jets and outflows are essential ingredients of the star formation process. A better understanding of this phenomenon is important in its own right as well as for many fundamental aspects of star formation.
   Jets and outflows associated with O-type protostars are rarely studied with observations reaching the close vicinity of the protostars.
   In this work, we report  high-resolution ALMA and VLBA observations to reveal a clear and consistent picture of an outflow associated with an O-type protostar candidate in the G26.50+0.28 region.
   These observations reveal, for the first time, a collimated jet located in the middle of the outflow cavity. The jet is found to be perpendicular to an elongated disk/toroid  and its velocity gradient. The collimated jet appears to show a small amplitude ($\alpha$$\approx$0$\ffad$06) counterclockwise  precession, when looking along the blueshifted jet axis from the strongest continuum source MM1, with a precession length of 0.22\,pc.
   The inclination of the jet is likely to be very low ($\approx$8$\ad$), which makes it a promising target to study its transverse morphologies and kinematics.
   However, no clear evidence of jet rotation is found in the ALMA and VLBA observations. The three-dimensional velocities of the water maser spots appear to show the same absolute speed with respect to different opening angles, suggesting the jet winds may be launched in a relatively small region. This favors the X-wind model, that is, jets are launched in a small area near the inner disk edge.}
   \maketitle

%
%-------------------------------------------------------------------

\section{Introduction}
\label{sec:intro}

Jets and outflows are natural outcomes in accreting astrophysical systems with rotation and magnetic fields \citep{1982MNRAS.199..883B} and are ubiquitous in star-forming regions
\citep[e.g.][]{2002A&A...383..892B, 2004A&A...426..503W, 2016ARA&A..54..491B}. Generally, the narrow and highly collimated ``jets'' are believed to arise through magnetohydrodynamic (MHD) processes in the rotating star-disk system. The less collimated and more massive ``outflows'' are believed to create shells of ambient gas swept up by the bowshock of the jet and/or a surrounding  wider-angle wind component \citep{2007prpl.conf..245A}.
Protostellar jets and outflows are essential elements of the star formation process.
They are proposed to remove the angular momentum from the young star and its disk, to determine the final core-to-star efficiency and envelope dissipation, to shift the core mass function to the stellar initial mass function, to counteract the dissipation of turbulence in clouds, and to impact planet formation through disk irradiation/shielding and MHD effects \citep[][]{2014prpl.conf..451F, 2016ARA&A..54..491B}.
Meanwhile, they also provide significant clues on the underlying star formation process, especially of deeply embedded and quickly evolving high-mass proto and young stellar objects, and provide a fossil record of the accretion history.
Therefore, the study of jets and outflows is essential in its own right as well as for a better understanding of many fundamental aspects of high-mass star formation.

How jets are launched and collimated is still under debate and the details of the interactions between jets and outflows remain unclear.
The long-standing X- or disk-wind problem, that jets are launched via magnetocentrifugal processes either from regions near the X-point  \citep[near the magnetospherical truncation on the disk; X-wind,][]{2000prpl.conf..789S} and/or from a wider area in the disk \citep[disk wind,][]{2000prpl.conf..759K},  is still with us.
One of the main reasons has been the inadequate angular resolution of current facilities needed to zoom into the launching and collimation zone \citep{2014prpl.conf..451F}.
High angular resolution observations reaching the close vicinity of the protostars or the base of the jet are expected to provide the most relevant information to deduce its launching and collimating processes.
In particular, the origin of jets from more massive protostars has rarely been investigated  so far \citep[e.g.][]{2014prpl.conf..451F, 2016ARA&A..54..491B}.
Outflows detected in high-mass ($\ga$8M$_{\odot}$) star-forming regions tended to be less collimated than those observed in low-mass star-forming regions \citep[e.g.][]{2002A&A...383..892B, 2003ApJ...598L.115T, 2007prpl.conf..245A, 2016A&A...585A..71M}, but there are also cases that show the contrary \citep[e.g.][]{2013ApJ...778...72F, 2020ApJ...905...25G, 2021MNRAS.507.4316B}.
%It remains controversial whether high-mass, especially O-type, protostars,  can launch collimated jets \citep[e.g.][]{2007prpl.conf..245A}.

Aperture synthesis imaging with interferometers, providing (sub)arcsecond angular resolution images and data cubes, is revolutionizing outflow studies.
Episodic ejection \citep[e.g.][]{2017A&A...597A.119T, 2015Natur.527...70P}, jet rotation \citep[e.g.][]{2017NatAs...1E.152L, 2020A&A...634L..12D, 2022A&A...668A..78D}, and collimated radial winds \citep[e.g.][]{2021ApJ...914L...1C, 2022ApJ...927L..27L} have been detected in observations of nearby star formation regions.
Very long baseline interferometry (VLBI) techniques, achieving even higher resolutions in the milli-arcsecond range, are another powerful tool to zoom into the part of the jet close to the launching point.
%Due to the sparse sampling in the UV space, VLBI observations can only trace the bright and compact masers.
Maser radiation is very sensitive to the excitation conditions and the geometry of the emitting regions, which makes it hard to derive the physical parameters from maser observations \citep{2007prpl.conf..197C}.
However, the three-dimensional (3D) velocities of the maser spots\footnote{A maser spot represents a compact emission peak in a specific velocity channel. Maser features arise from the same region but extend over several contiguous velocity channels.} derived from multiple epoch VLBI observations, allow us to study the kinematics at the base of the jet.
VLBI observations %filter out most of the diffuse emission,
only address the kinematics of the tiny regions encompassed by the masers \citep{1992ARA&A..30...75E}.
The selective kinematical information, albeit sparse, %eliminates the contamination of the emission from the ambient gas,
allows us to unambiguously interpret the kinematical distribution of the jet winds.
In practice, the interpretation of maser emission might be complicated and ambiguous, and sometimes the same maser species have been claimed to trace disks and outflows simultaneously.
To achieve a comprehensive understanding of the jet properties, it is necessary to complement VLBI observations with high angular resolution imaging of thermally exited gas \citep[e.g.][]{2022A&A...659A..81B}.

Located at a maser parallax distance of 6.3$^{+0.5}_{-0.4}$ kpc \citep{2021ApJS..253....1X}, G26.50+0.28 (also known as IRAS\,18360-0537) has a far-Infrared luminosity of 1.2$\times$10$^5$\,L$_{\odot}$ (Molinari et al. 1996).
The continuum map obtained with the Submillimeter Array (SMA) at 1.3 mm reveals two unresolved condensations, MM1 and MM2, in the source, while SMA maps of the $^{13}$CO (2-1) and SiO (5-4) line emission show a northeast-southwest wide-angle bipolar outflow centered at MM1 \citep{2012ApJ...756..170Q}. Images of CH$_{3}$OH and CH$_{3}$CN emission reveal a ``disk/toroid''-like rotational structure with the velocity gradient perpendicular to the outflow axis. Furthermore, CN spectra, also from the SMA, present typical inverse P-Cygni profiles that demonstrate infall motions of about 1.5$\times$10$^{-3}$ M$_{\rm \odot}$ yr$^{-1}$ (Qiu et al. 2012). Given its luminosity, high accretion rate, and massive gas reservoir in the envelope, \citet{2012ApJ...756..170Q} suggested that MM1 would eventually form an O-type star.
In its environment, G26.50+0.28 is one of the few isolated O-type targets, neighboring only a quiescent dense core, MM2.

In this paper, we focus on the study of the jet/outflow system in G26.50+0.28 by using three-configuration Atacama Large Millimeter/submillimeter Array (ALMA) data and fifteen epoch Very Long Baseline Array (VLBA) water maser data, with resolutions of $\approx$ 0.2$\as$ (1260\,au) and $\approx$0.002$\as$ (13\,au), respectively, to achieve a comprehensive understanding of the jet properties in G26.50+0.28.

%--------------------------------------------------------------------
\section{Observations}
\label{obs}
\subsection{ALMA data}

The archival ALMA data of G26.50+0.28, including those of the 1.3\,mm continuum and the typical outflow tracers $^{13}$CO(2-1) (220.399\,GHz) and SiO(5-4) (217.105\,GHz), were observed in Cycle 6 as part of  the ALMA Large project ALMAGAL: ALMA Evolutionary study of High Mass Protocluster Formation in the Galaxy (2019.1.00195.L, PI: Sergio Molinari).
These data are used to study the large-scale structures of the outflow and also allow us to understand the origin of the water masers.
The ALMA observations were carried out in Cycle 6 with the Atacama Compact Array (ACA) (with baselines in the range 8.85--48.95\,m) and 12-m array in C-4 (with baselines the range 15.06--783.54\,m) and C-6 (with baselines in the range 15.06--2516.86\,m) configurations.  The on source times for 7m, C-4, and C-6 configuration observations are 100s, 54s, and 120s, respectively.
These observations used J1924-2914 as the flux and band pass calibrators and J1832-1035 as the gain calibrator.
See Table 1 for the details of the setups.

The data were calibrated and imaged using the Common Astronomy Software Applications (CASA) package \citep{2007ASPC..376..127M} version
5.6.1.8 for the 7m and C-4 observations and 6.1.1 for the C-6 observations.
The 7\,m and two configuration 12\,m data were first processed via the standard ALMA reduction pipeline separately (pipeline version 42866M for the 7m observations, 42866 for the C-4 observations, and 2020.1.0.40 for the C-6 observations) and then combined following standard procedures.
We used the \textsf{multiscale} deconvolver \citep{2008ISTSP...2..793C} with scales of zero (delta-function) and 6, 10, and 20 times the pixel size, which are about 1, 2, and 4 times the beam size, to recover both compact and extended emission.
The continuum image was produced using the multi-frequency synthesis \citep[mfs; ][]{1990MNRAS.246..490C} algorithm  over a frequency range identified to lack line emission.
For the spectroscopic data cubes, we first used the CASA task \textsf{uvcontsub} with the polynomial order of unity to subtract the continuum emission in the $uv$ domain and then cleaned the channels with a width of 1\,$\kms$ in the range of 70-140 $\kms$.
For all the data products, we used  \textsf{Briggs} cleaning and set the robustness parameter \textsf{robust} to the default value of 0.5, which provides a good compromise between noise and resolution.  The continuum image reaches a 1$\sigma$ noise level of 0.3 m$\Jypb$ in a synthesized beam of 0$\ffas$30$\times$0$\ffas$17 with P.A. $\approx$-65$\ad$.
The synthesized beams for $^{13}$CO\,(2-1) and SiO\,(5-4) with $\approx$ 0$\ffas$36$\times$0$\ffas$24 and PA $\approx$-57$\ad$ are similar.
The typical 1$\sigma$ noise level of the two line images is $\la$ 5.0\,m$\Jypb$ in a 1\,$\kms$-wide channel.
%
%ALMA observations, with 2--3 times better resolution than previous SMA data \citep{2012ApJ...756..170Q}, have greatly revolutionized our understanding of the jet/outflow in G26.50+0.28 (see Section \ref{sec:alma}).
%
In this paper, ALMA maps represent the combined ACA, C-4, and C-6 data.

\begin{table}
\caption{Summary of the spectral setups of the ALMA observations.}
\label{tab:almaobs}
\centering
\begin{tabular}{cccc}     % 7 columns
\hline\hline
 SPW  & Freq. Cove. & Channel Width   &  Channels \\
      &  (GHz)  &  (kHz)  & \\
 (1) & (2)  & (3) & (4)     \\
 \hline\hline
\multicolumn{4}{c}{7m ACA} \\
\multicolumn{4}{c}{Obs. Date: Oct. 22, 30 2019, Mar. 07 2020} \\
\hline
1  &  216.8-218.8  & 488.281  &  4096 \\
2  &  218.8-220.8  & 488.281  &  4096 \\
3  &  217.9-218.5  & 122.281  &  4096 \\
4  &  220.3-220.8  & 122.281  &  4096 \\
 \hline
\multicolumn{4}{c}{12m C-4} \\
\multicolumn{4}{c}{Obs. Date: Oct. 12, 13, 2019} \\
\hline
1  &  217.0-218.8  & 488.281  &  3840 \\
2  &  219.0-220.8  & 488.281  &  3840 \\
3  &  217.8-218.2  & 122.281  &  3840 \\
4  &  220.3-220.8  & 122.281  &  3840 \\
\hline
\multicolumn{4}{c}{12m C-6} \\
\multicolumn{4}{c}{Obs. Date: May 27, June 6, 7, 9 2021} \\
\hline
1  &  217.0-218.8  & 488.281  &  3840 \\
2  &  219.0-220.8  & 488.281  &  3840 \\
3  &  217.8-218.2  & 122.281  &  3840 \\
4  &  220.3-220.8  & 122.281  &  3840 \\
 \hline
\end{tabular}
\tablefoot{Columns (1) -- (4) are spectral windows, ranges of rest frequencies covered by the spectral windows, channel widths, and channel numbers.}
\end{table}

\subsection{VLBA data}

Fifteen epoch VLBA water maser (22.23508\,GHz) observations of G26.50+0.28 were conducted from 2015 March 25 to 2016 May 7 under program BR210, which is part of the VLBA Key Science project, the Bar And Spiral Structure Legacy (BeSSeL) Survey\footnote{http://bessel.vlbi-astrometry.org}.
The intensive observations spanning about 13 months are important for constraining the time-dependent effects of water masers themselves.
A total of 2000 8\,kHz ($\approx$0.11\,$\kms$) channels centered at V$_{\rm LSR}$ = 104\,$\kms$ (Channel 1001) were used.
The data were calibrated and imaged using the Astronomical Image Processing System (AIPS) scripts written in ParselTongue \citep{2006ASPC..351..497K}.
Calibration procedures were conducted following the procedures of the BeSSeL survey \citep{2009ApJ...700..137R}.
In this case,  masers are much stronger than the background sources. Therefore, inverse phase-referening was applied, where the fringe fit in all the epochs was performed with a strong and point-like water maser spot identified at V$_{\rm LSR}$ = 104.11 $\kms$ (channel 1000) and (R.A., DEC)$_{\rm J2000}$ = (18:38:40.1707, -05:35:42.7749). Then the calibration was transferred to the continuum background sources and the other spectral channels.
The synthesized beam size varies from epoch to epoch, but is typically about 2.0 $\times$ 1.0 mas (13$\times$6\,au) (see Table \ref{tab:epoch}). We fit two-dimensional Gaussian brightness distributions to each identified maser spot to obtain its position using the AIPS task \textsf{SAD}, resulting in $\ga$ 200 water maser spots in each of the 15 epochs (see Table \ref{tab:epoch}); refer to \citet{2021ApJS..253....1X} for detailed information on the observations.

We should note that the maser spots are mainly from $\approx$ 30 maser features (a group of maser spots being continuous in velocity)  (see Table \ref{tab:epoch}).
The maser features and their proper motions are shown in Fig. \ref{fig:feature}, which are consistent with the results derived from the maser spots (see Section \ref{sec:pm}).
However, we find that the standard deviation of the amplitudes of the proper motions derived from maser features in the two maser clusters (see Section \ref{sec:pm}) is larger than that derived from maser spots.
This may be the case because the position of a feature was estimated based on the flux-weighted positions of the spots within the feature, which could potentially introduce additional errors in the position determination. For instance, features can comprise varying numbers of maser spots at different epochs due to the rapid intensity variability of water masers.
%
%This may be the case because the estimation of the flux-weighted positions of features from spots introduce additional position errors. Furthermore, features consist of different amounts of maser spots at different epochs due to the rapid intensity variability of water masers.
%
Therefore, in this work, we use the multi-epoch water maser spots to conduct internal motion studies, which has been proved to be capable of determining the proper motions of the outflowing H$_2$O masers in star-forming regions \citep[e.g.][]{2011MNRAS.410..627T,2013ApJ...775...79Z}.

\begin{table*}
\caption{Epochs of the observations and detected H$_{2}$O maser spot numbers of the 15 VLBA observations.}
\label{tab:epoch}
\centering
\begin{tabular}{ccccccc}     % 7 columns
\hline\hline
 Epoch  & Obs. Date  &    beam(mas)    &   P.A.     & N$_{\rm s}$   &   N$_{\rm f}$  & jet/wind \\
 (1) & (2)  & (3)    & (4)   & (5)  &  (6)  &  (7) \\
 \hline\hline
E2   & Mar. 25 2015  &  1.80$\times$1.10 &  8$\ad$   &  262  &  35   &   0.535 \\
E3   & Apr. 19 2015  &  2.01$\times$1.20 &  22$\ad$   &  264  &  28   &   0.581 \\
E4   & May  08 2015  &  1.65$\times$1.13 &  42$\ad$   &  252  &  27   &   0.605 \\
E5   & Aug. 19 2015  &  1.72$\times$0.81 &  -1$\ad$   &  169  &  24   &   0.598 \\
E6   & Sep. 07 2015  &  2.65$\times$1.12 &  -5$\ad$   &  185  &  21   &   0.603 \\
E7   & Sep. 18 2015  &  2.14$\times$1.00 &  -3$\ad$   &  208  &  23   &   0.613 \\
E8   & Oct. 01 2015  &  1.67$\times$1.02 &  -1$\ad$   &  234  &  28   &   0.606 \\
E9   & Oct. 12 2015  &  1.70$\times$0.73 &  -7$\ad$   &  215  &  27   &   0.598 \\
E10  & Oct. 26 2015  &  1.74$\times$0.68 &  -8$\ad$   &  227  &  27   &   0.587 \\
E11  & Nov. 05 2015  &  1.83$\times$1.09 &  20$\ad$   &  255  &  31   &   0.577 \\
E12  & Nov. 17 2015  &  1.72$\times$1.06 &  13$\ad$   &  230  &  30   &   0.579 \\
E13  & Feb. 18 2016  &  2.44$\times$0.60 &  -13$\ad$   &  251  &  25   &   0.583 \\
E14  & Mar. 15 2016  &  1.67$\times$1.00 &  25$\ad$   &  234  &  24   &   0.589 \\
E15  & Apr. 11 2016  &  1.49$\times$0.75 &  -8$\ad$   &  253  &  25   &   0.594 \\
E16  & May  07 2016  &  2.41$\times$0.86 &  -12$\ad$   &  209  &  24   &   0.599 \\
 \hline
\end{tabular}
\tablefoot{Column 1: epoch name. Column 2: Observing dates. Columns 3 and 4: The size and P.A. of the beam. Columns 5 and 6: The total number of maser spots and features detected. Column 7: The ratio between the total flux of the maser spots associated with the dense jet, i.e. maser spots in MC-1 and MC-2 (see Fig. \ref{fig:mc12}), and the total flux of the ones in the blue lobe (see the right panel of Fig. \ref{fig:maserDis}). }
\end{table*}

\subsection{Herschel data}

Far-infrared data were observed by the Herschel Space Observatory which was built and operated by the European Space Agency (ESA) \citep{2010A&A...518L...1P}.
Herschel SPIRE 250$\mu$m data of G26.50+0.28 were derived from the Herschel High Level Images (HHLI)\footnote{https://irsa.ipac.caltech.edu/data/Herschel/HHLI}.
The data have been processed to Level 2.5 through the Standard Product Generation pipeline, version 14.0.
The angular resolution of the SPIRE 250 $\mu$m data is 18$\as$ \citep{2010A&A...518L...3G}. The SPIRE image data is calibrated in $\Jypb$.

\section{Large-scale structures reveled by ALMA observations}
\label{sec:alma}

\subsection{1.3\,mm continuum}
\label{sec:disk}

\begin{figure*}
    \centering
    \includegraphics[width=0.49\hsize]{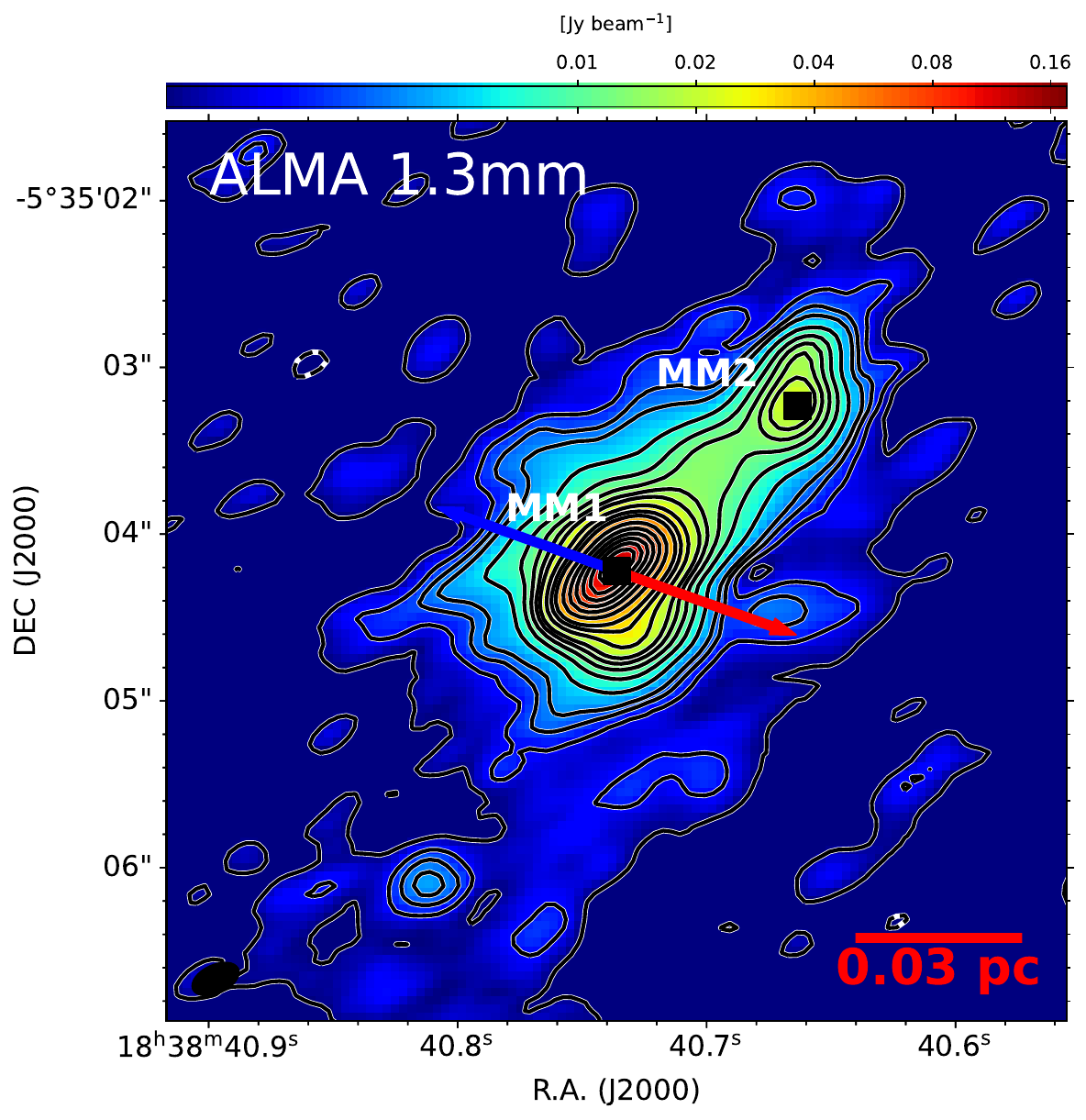}
    \includegraphics[width=0.49\hsize]{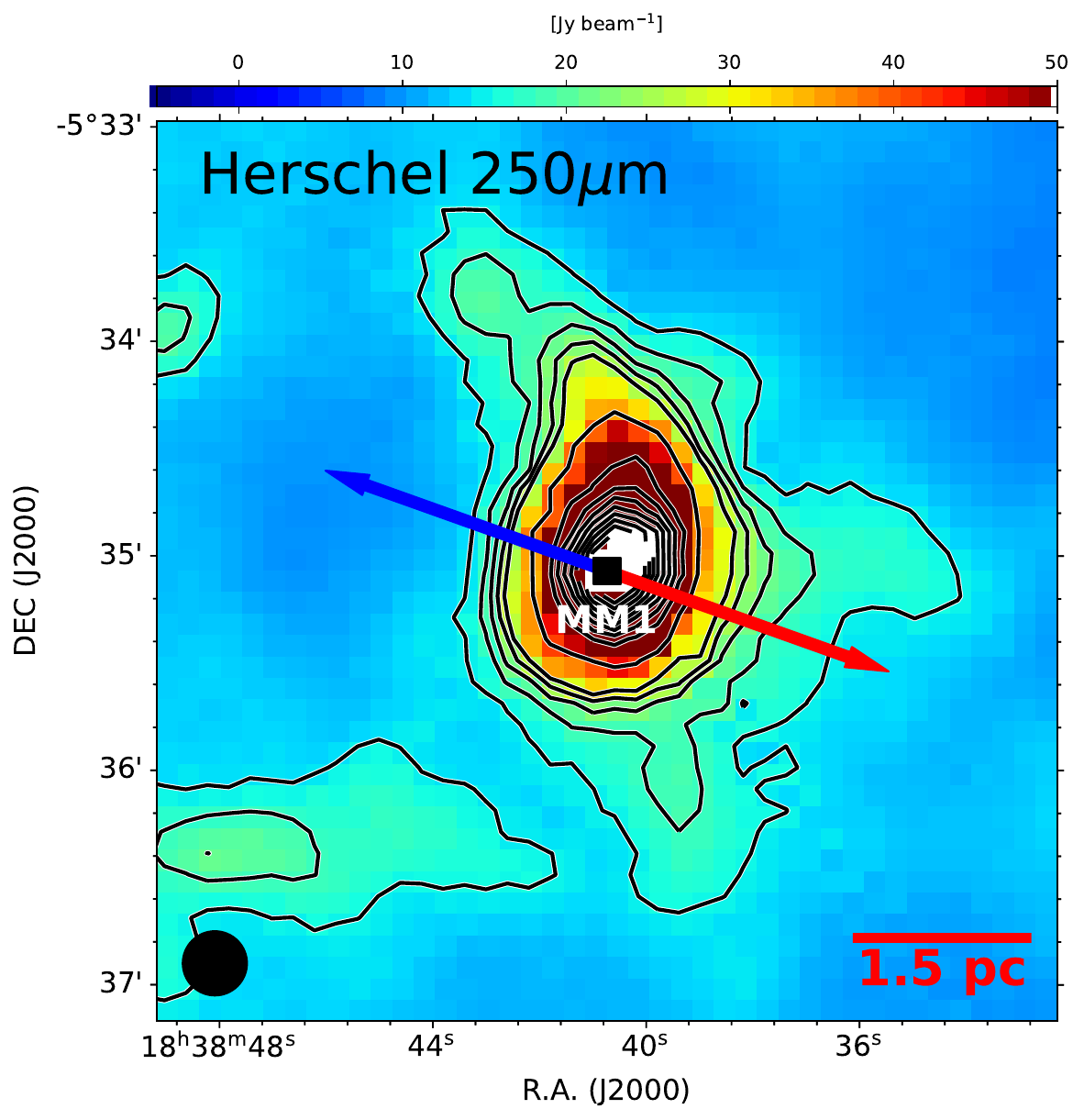}
    \caption{ALMA 1.3mm continuum (left panel) and Herschel 250 $\mu$m (right panel) images of G26.50+0.28. The contour levels in the left and right panels are set to [-3, 3, 6, 9, 12, 20, 30, 40, 50, 60, 80, 100, 120, 160, 200, ..., 360] $\times$ 0.3\,m$\Jypb$ and [15, 18 ....30, 40, 70, ..., 310] m$\Jypb$, respectively. Dashed contour lines represent negative values.
    In each panel, a filled ellipse in the lower left shows the beam; The blue and red arrows indicate the directions of the jet; The filled squares denote the peak positions of the two cores which are labeled as MM1 and MM2.}
    %\textbf{The position error can be roughly estimated as 0.5$\times$ $\theta_{beam}$/(S/R)$\approx$.} }
    \label{fig:cont}
    \end{figure*}

The 1.3\,mm continuum radiation is completely dominated by dust emission, due to the faintness of the free-free emission \citep{2012ApJ...756..170Q}.
In the left panel of Fig.\ref{fig:cont}, two dense cores, MM1 and MM2, are identified in the ALMA 1.3\,mm continuum image, which is consistent with the previous SMA observations.
%
%Compared with the SMA observations,
The ALMA image also shows diffuse emission connecting MM1 and MM2, while, on larger scales, more diffuse emission arises from a filamentary structure roughly extending in a  north-south direction as seen in the right panel of Fig. \ref{fig:cont}.
In the high-resolution ALMA image, the MM1 core is slighly resolved, enabling us to see the elongated morphology of MM1 for the first time. This is also true for the ALMA 1.3\,mm continuum image restored with a 0$\ffas$3 circular beam (see Fig. \ref{fig:cont_bmc}).
Meanwhile, the elongation is perpendicular to the putative collimated jet (see below), making the target a potential jet-disk system.
Furthermore, the inclination of the jet to the plane of the sky should be smaller (due to the disk thickness) than $\arctan(b/a)$ $\approx$32$\ad$, assuming a circular disk/toroid, where the deconvolved major and minor diameters $a$ and $b$ are 0$\ffas$65 (4088 au) and 0$\ffas$41 (2608 au), respectively.
Vice versa, the scale height H of the disk/toroid can be also roughly estimated, complemented by the relatively robust jet inclination  of 8$\ad$ with respect to the plane of the sky (see Section \ref{sec:vlba}).
Assuming the jet is perpendicular to the disk, then $H = (b-a \times$ $\sin$(8$\ad$))/$\cos(8$\ad$)$ $\approx$ 0$\ffas$33 (2060 au), which is about half of the major diameter of the disk/toroid.
The scale height is also related to the mass of the central star \citep[e.g.][]{2014A&A...566A..73C},
\begin{equation}
\label{height}
  H =\frac{FWHM}{\sqrt{8\ln(2)}}\sqrt{\frac{R_{disk}^{3}}{\rm G \it M_{\star}}},
\end{equation}
where the FWHM linewidth is set to 6 $\kms$ which is the value %median FWHM linewidth
found for the 217.3\,GHz $\nu_{t}$=1  6(1,5)--7(2,6) line of the typical disk tracer CH$_{3}$OH  from the ALMA data. The radius of the disk, $R_{\rm disk} = a/2$, is 0$\ffas$325 (see above).
Then the derived mass of the central star $M_{\star}$ is about 15 M$_{\odot}$, which is consistent with the mass derived by \citet{2012ApJ...756..170Q} of 12 M$_{\odot}$.

\begin{figure*}
    \centering
    \includegraphics[width=0.49\hsize]{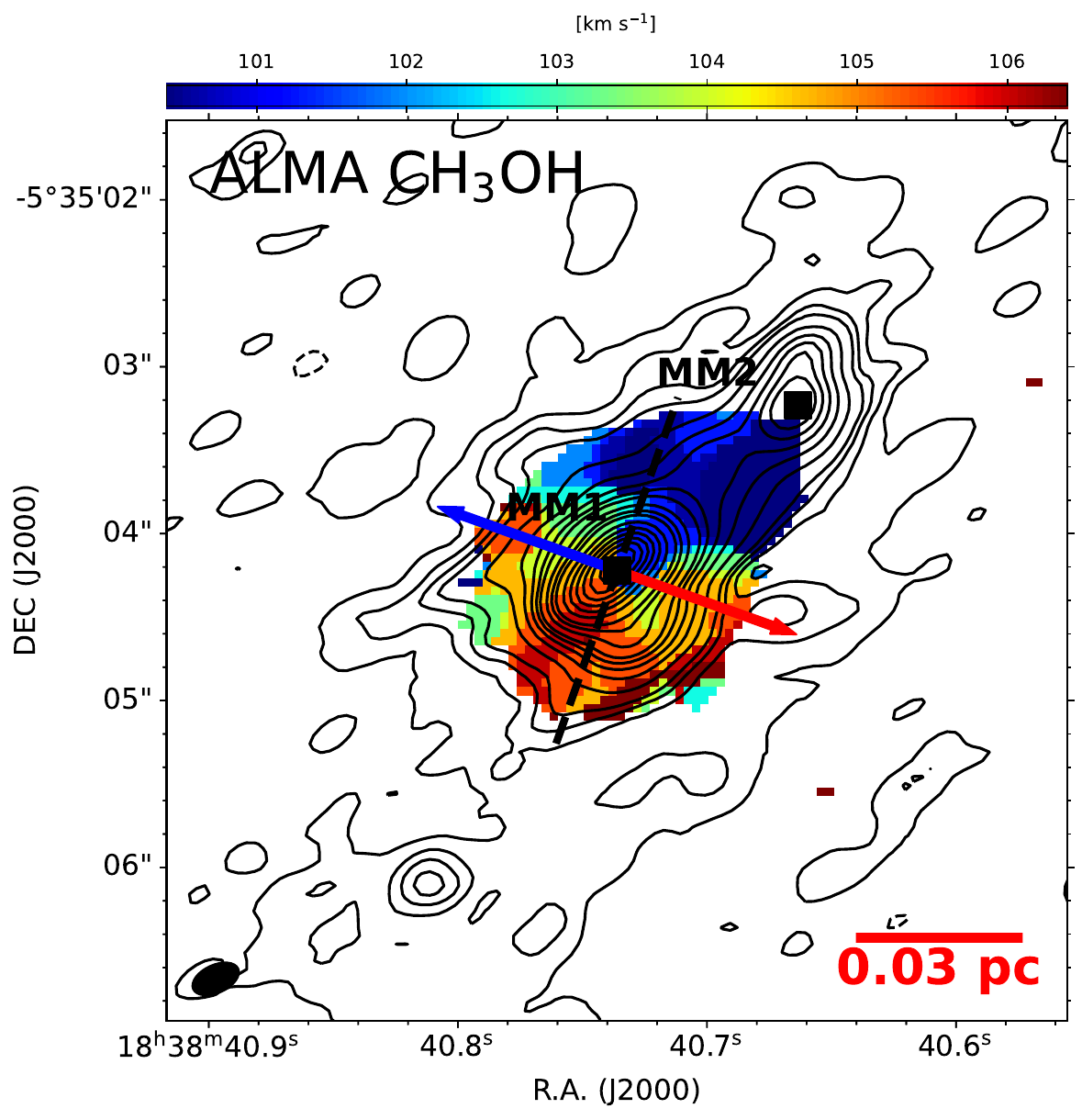}
    \includegraphics[width=0.49\hsize]{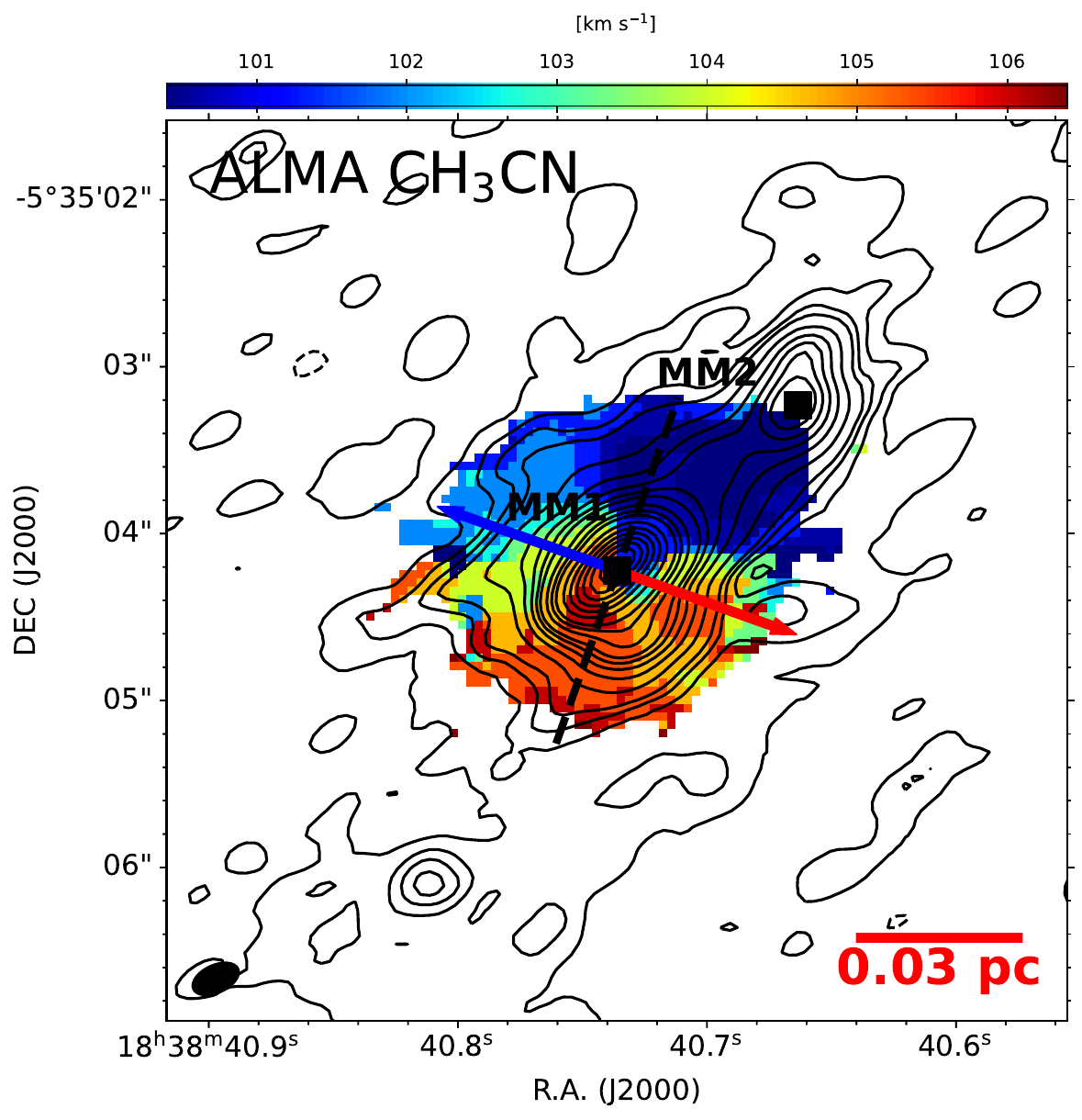}
    \includegraphics[width=0.49\hsize]{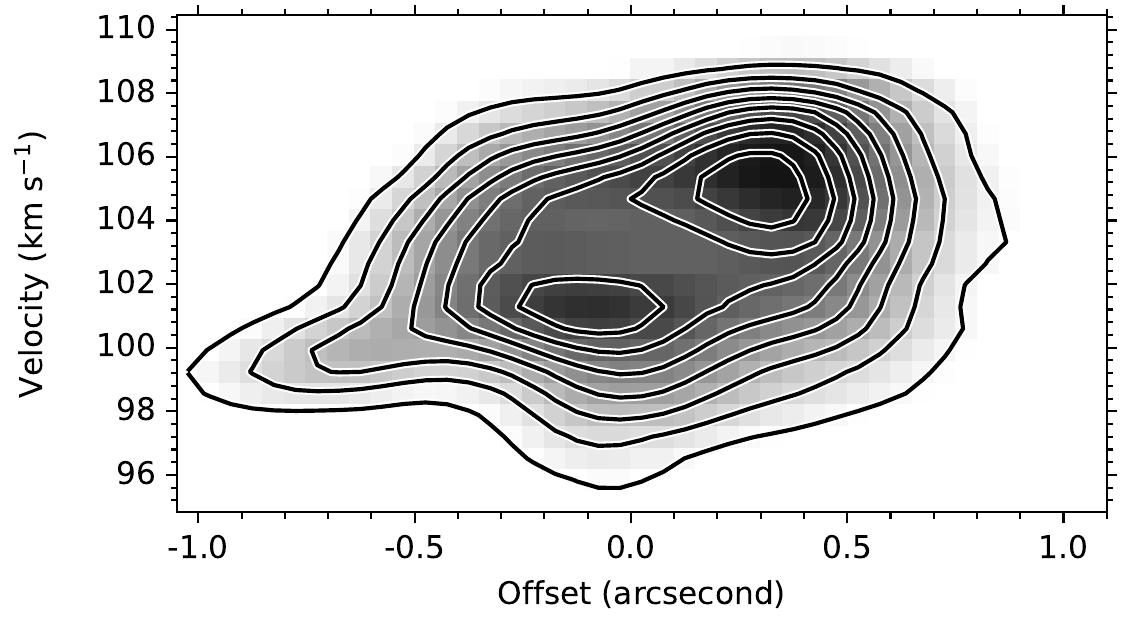}
    \includegraphics[width=0.49\hsize]{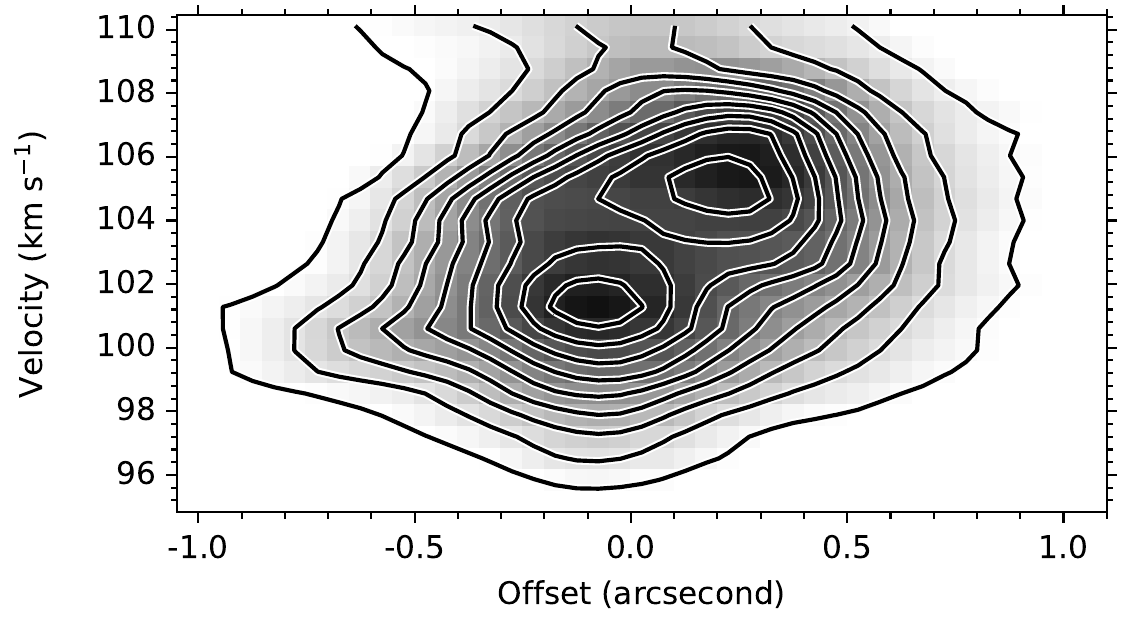}
    \caption{ALMA velocity of the maximum value of the spectrum (peak velocity) images of CH$_{3}$OH (6$_{1,5}$-7$_{2,6}$ $\nu_{t}$=1) (top left panel) and CH$_{3}$CN (12$_{7}$-11$_{7}$ $\nu_{t}$=1) (top right panel)). A filled ellipse in the lower left of each panel shows the FWHM beam sizes.
    The contours overlaid in the two panels are the same as those in the left panel of Fig. \ref{fig:cont}. The blue and red arrows indicate the symmetry axis  of the jet. The filled squares denote the peak positions of the two cores which are labeled as MM1 and MM2. A dashed line in each of the top panels illustrates the P-V cut perpendicular to the jet. The lower panels show the P-V diagrams of CH$_{3}$OH (6$_{1,5}$-7$_{2,6}$ $\nu_{t}$=1) (left) and CH$_{3}$CN (12$_{7}$-11$_{7}$ $\nu_{t}$=1) (right). The contours all start from 28 m$\Jypb$ and increase in steps of 28 m$\Jypb$.
    }
    \label{fig:cont-vel}
    \end{figure*}

The velocity gradients, found with the typical disk tracers CH$_{3}$OH (6$_{1,5}$-7$_{2,6}$ $\nu_{t}$=1) and CH$_{3}$CN (12$_{7}$-11$_{7}$ $\nu_{t}$=1) \citep[][]{2016A&ARv..24....6B}, are roughly perpendicular to the jet (see Fig. \ref{fig:cont-vel}).
The velocity gradient is also generally consistent with that revealed by previous SMA data.
The higher resolution ALMA data seem to present an even thinner disk with a higher velocity gradient in MM1 (see Fig.\ref{fig:cont-vel}).
%
%The higher resolution ALMA data further present a more complicated velocity field than a uniform velocity gradient. An even thinner disk with a velocity gradient is likely found in MM1.
%
However, the resolution of the ALMA data is not high enough to fully resolve it.
Being also beyond the scope of this work, it will be followed up in a future contribution based on higher resolution data.

\subsection{$^{13}$CO (2-1) and SiO (5-4) maps}
\label{sec:CO_SiO}
\begin{figure*}
    \centering
    \includegraphics[width=0.49\hsize]{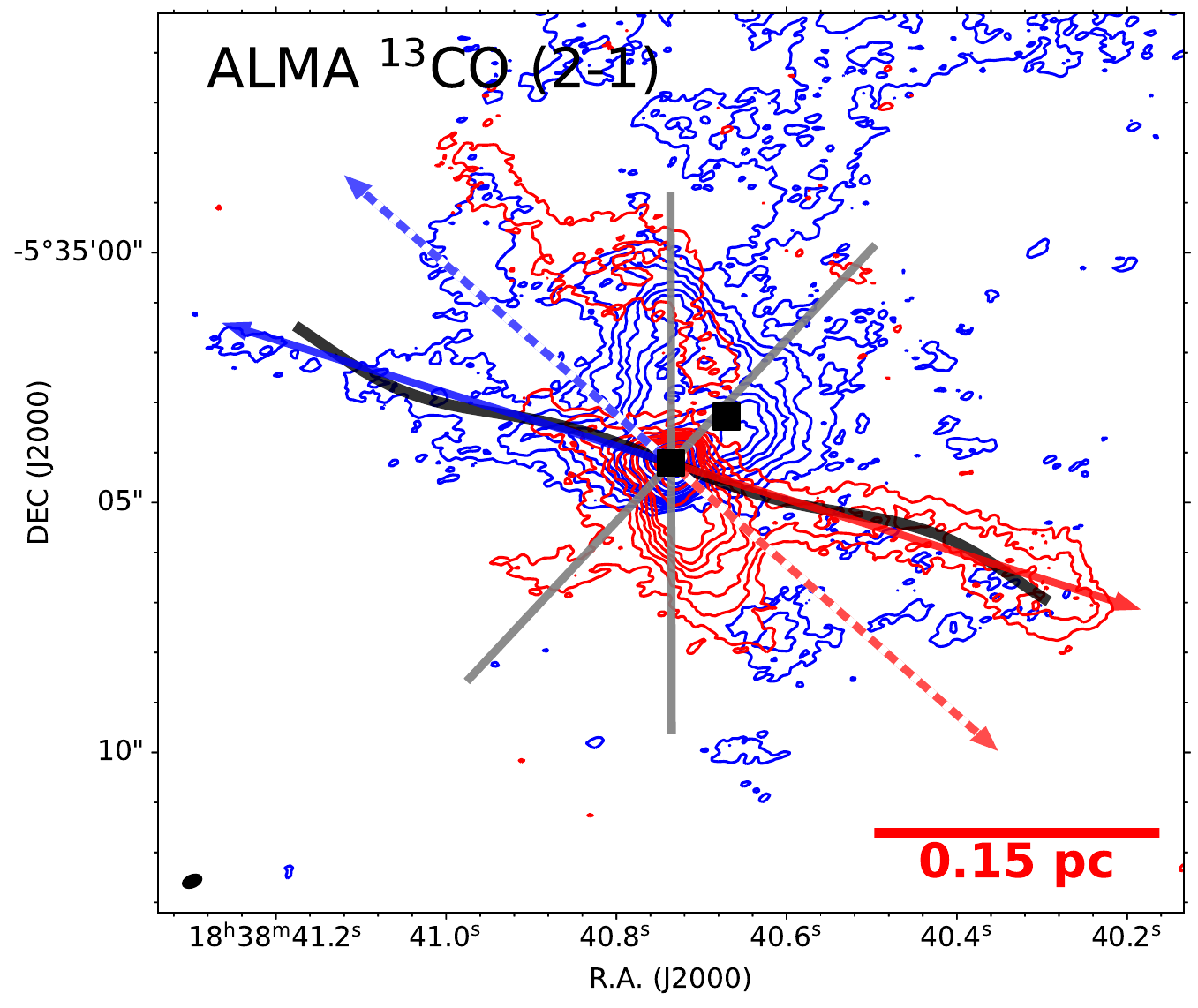}
    \includegraphics[width=0.49\hsize]{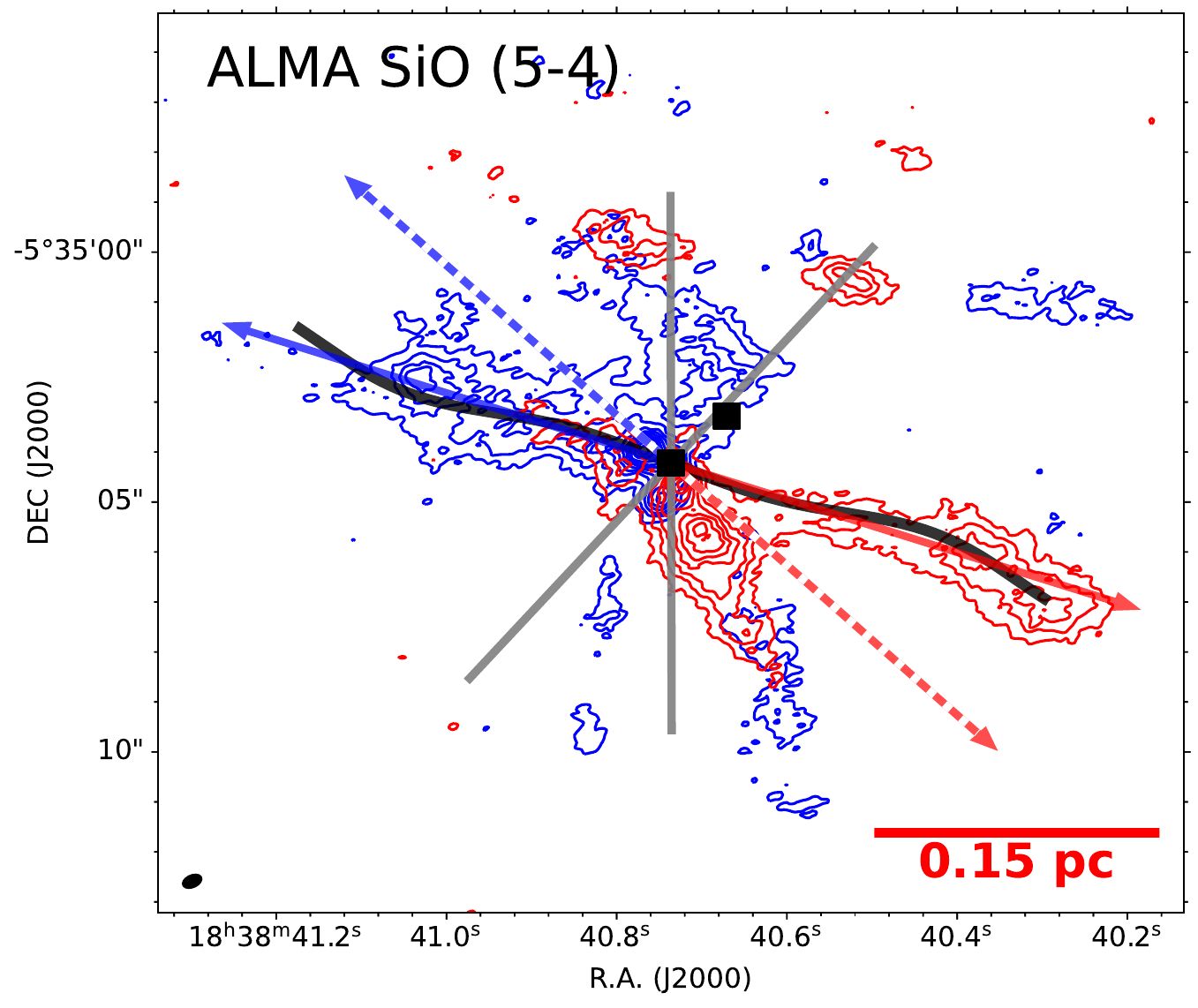}
    \caption{ALMA $^{13}$CO (2-1) (left panel) and SiO(5-4) (right panel) images of G26.50+0.28.
    The blue contours in each panel are integrated from 70  to 103 $\kms$ and the red ones are integrated from 105 to 138 $\kms$.
    The contour levels of $^{13}$CO (2-1) are set to [4, 8, 12, 16, 20, 24, 28, 40, 52, 64, 76, 88, 100]$\times$ 28.5 m$\Jypb$ $\kms$.
    The SiO\,(5-4) levels are set to [4, 8, 12, 16, 20, 24, 28, 32,  40, 48]$\times$ 28.7 m$\Jypb$ $\kms$.
    The blue and red arrows in the two panels indicate the direction of the jet.
    The dashed blue and red  arrows denote the axis of the biconical outflow determined by \citet{2012ApJ...756..170Q}.
    The solid gray lines denote the shells of the wide-angle outflow with opening angle of about 140$\ad$.
    A filled ellipse in the lower left of each panel shows the beams. The filled squares denote the peak positions of MM1 and MM2. The black curved lines denote the loci of the precessing jet.
    }
    \label{fig:CO_SiO}
    \end{figure*}

Fig. \ref{fig:CO_SiO} presents the velocity-integrated maps of two typical jet/outflow tracers, the $^{13}$CO (2-1) (left panel) and the SiO (5-4) lines (right panel).
A collimated jet-like structure centered at MM1, denoted by the solid blue and red arrows, is presented in both panels of Fig. \ref{fig:CO_SiO}, which is more prominent in the SiO(5-4) map.
A redshifted clump shown at the base of the blue lobe may be produced by jet precession (see Section \ref{sec:jet-precession}).
%%%%   KMM: WHAT DOES  "exceed the inclination of the jet" MEAN?
As mentioned  in Section \ref{sec:disk}, the axis of the jet is perpendicular to the elongation of the disk/toroid and its velocity gradient.
Meanwhile, this collimated jet is consistent with the distribution of the VLBA water masers (see Section \ref{sec:vlba}).

The remaining blueshifted and redshifted  $^{13}$CO (2-1) emission has a complex distribution, forming shell-like structures on either side of the collimated jet (except for the  relatively diffuse emission in the northern part of the $^{13}$CO (2-1) map), which may be enhanced by limb-brightening.
%
%\textbf{We also present the low velocity emission of $^{13}$CO (2-1) in Fig. B3, which shows the more clearly shell structures, since the entrain shell by the jet/wind is usually more obvious in the low velocity emission (e.g. Plunkett et al. 2015).}
%
%
The shells have an opening angle of about 140$\ad$, denoted by the gray lines  in Fig. \ref{fig:CO_SiO}.
The mixture of blueshifted and redshifted emissions may be explained by the combined effect of wide-angle winds and the small inclination of the jet/outflow.
The two shells in the roughly north-south direction show relatively  strong $^{13}$CO (2-1) and  SiO(5-4) emissions in the two panels of Fig. \ref{fig:CO_SiO}.

We note: (1) the two sided shells do not show point- or mirror-symmetry relative to MM1 or MM2, especially the northern shell is bending by about 90$\ad$  (see the right panel of Fig. \ref{fig:CO_SiO}), which may be not commonly seen in jets. Actually, the overall morphology of the shells roughly follows the large-scale north-south filamentary structure seen in the right panel of Fig. \ref{fig:cont}. Therefore it is likely caused by the interactions between the wide-angle winds with the dense filamentary structure;
(2) most of the water masers are exclusively associated with the collimated jet and barely with the shells (see Section \ref{sec:vlba}).
Therefore, it is likely that Fig. \ref{fig:CO_SiO} presents a collimated dense jet with wide-angle wind components. The wide-angle winds are interacting with the large-scale filamentary structure, enhancing the $^{13}$CO (2-1) and SiO(5-4) emission in the north-south shells.
%\textbf{However, the possibility that an additional jet is hiding in the relatively strong north-south shells cannot be excluded.}
%

We should note that the possibility that an additional jet is hiding in the relatively strong north-south shells cannot be excluded.
Meanwhile, as mentioned in the introduction, based on the SMA data, \citet{2012ApJ...756..170Q} argued that there is a wide-angle northeast-southwest bipolar outflow centered at MM1. The axis of the outflow is denoted by the dashed red and blue arrows in Fig. \ref{fig:CO_SiO} with P.A. $\approx$ 45$\ad$.
However, our high resolution and sensitivity ALMA data have further resolved the outflow and detected weaker emission of the shells.
Complemented by the water maser distribution and its kinematics (see Section \ref{sec:vlba}), we derive a different interpretation, i.e. a collimated jet and wide-angle wind scenario.
We also present the low velocity emission of $^{13}$CO (2-1) in Fig. \ref{fig:COlv}, which shows more clearly the shell structures, since entrained shells by the jet/wind are usually more obvious in tracers of the low velocity emission (e.g. Plunkett et al. 2015).

\subsection{Jet precession}
\label{sec:jet-precession}
We can see from Fig. \ref{fig:CO_SiO} that the two lobes of the collimated jet seem to present point-symmetric (S-shaped) wiggles relative to MM1, suggesting precession.
In order to model it, we used a simple sinusoidal equation \citep{2009ApJ...698..184W}, adapted from \citet{1996AJ....112.2086E} to fit the SiO(5-4) emission,

\begin{equation}
%\begin{split}
\label{jet_precession}
\left( \begin{matrix} x \\ y \end{matrix} \right)=  \left( \begin{matrix} \cos(\psi) & -\sin(\psi) \\ \sin(\psi)& \cos(\psi) \end{matrix} \right) \times  \left( \begin{matrix} \alpha l \sin(2\pi l /\lambda + \phi_{0} ) \\ l \times cos(i) \end{matrix} \right),
%\end{split}
\end{equation}
where $x$ and $y$ are the coordinates relative to MM1. $\alpha$  and $\lambda$ are the amplitude and wave scale of the precession. $\phi_{0}$ is the initial phase at the location of MM1. $\psi$ is the position angle of the jet symmetry axis in the plane of the sky.  $i$ is the inclination angle of the jet to the plane of the sky.
The fitting results are summarized in Table \ref{tab:jet-precession} and illustrated in Fig. \ref{fig:CO_SiO} by the solid black curves.
We can see that the fitting roughly recovers the wiggles of the two lobes. However, we should note that the amplitude of the jet wiggles is small. Therefore the fitting  could not be well confined, which may be the reason for the relative poor fitting of the blueshifted lobe.

\begin{figure*}
    \centering
    \includegraphics[width=0.49\hsize]{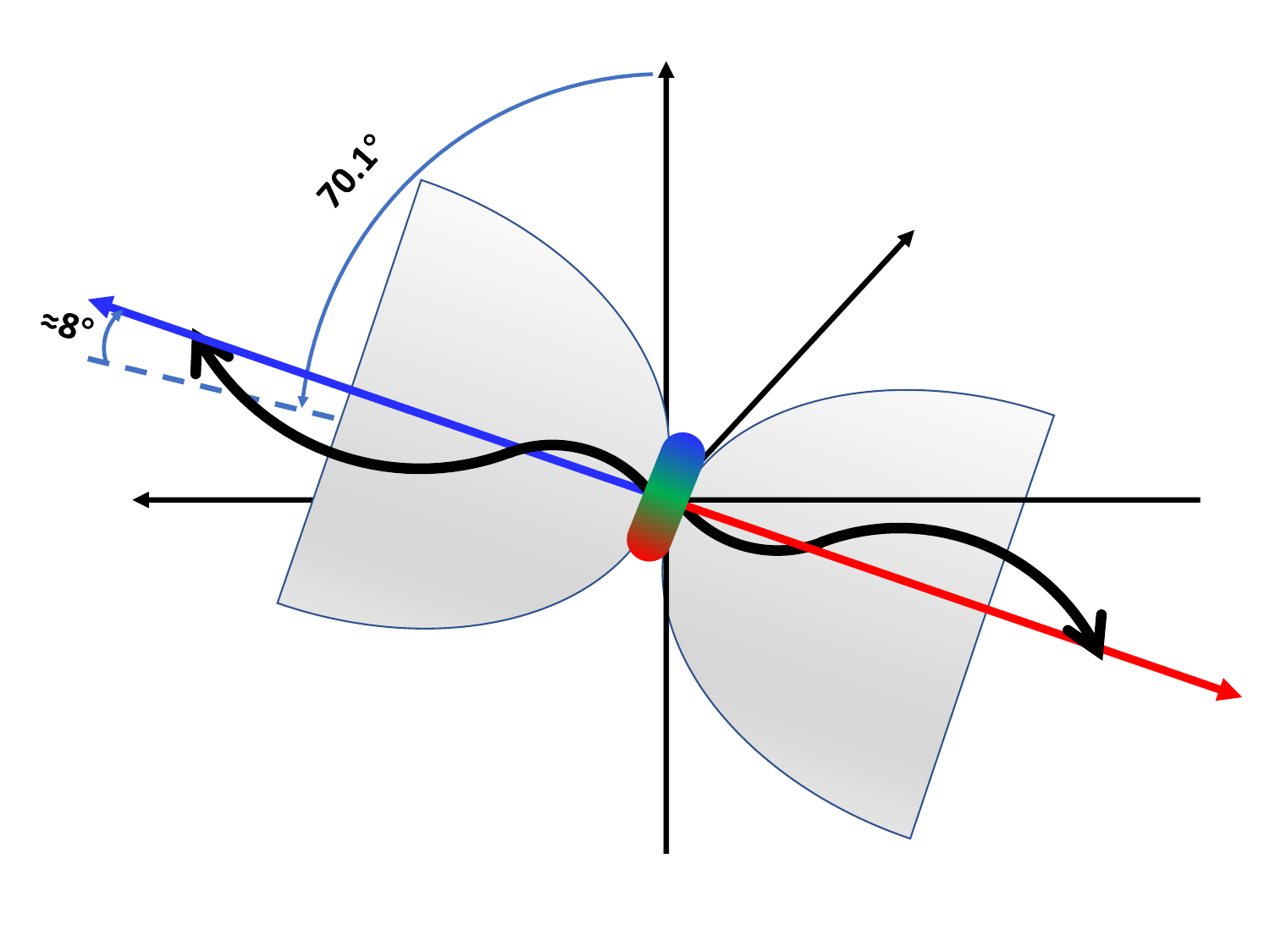}
    \caption{Sketch of the jet/outflow in G26.50+0.28.}
    \label{fig:jet-sch}
    \end{figure*}

The derived precession length $\lambda$ is 7$\ffas$2, which is 0.22\,pc at a distance of 6.3\,kpc.
%
%Assuming a jet velocity\footnote{Here we do not use the jet velocity derived by the water maser spots in Section \ref{sec:pm}. The water maser velocity may be generally smaller than the possible jet velocity, due to entrainment of the pre-shock material \citep[e.g.][]{1993ApJ...414..230M, 2016A&A...585A..71M}.} of $\sim$ 200\,$\kms$ \citep[e.g.][]{2018A&ARv..26....3A}, the precession period should be at the order of 1000 years.
%
Assuming a jet velocity of $\sim$ 30\,$\kms$ (see Section \ref{sec:pm}), the precession period should be at the order of 7000 years. We should note that the period may be overestimated, because the water maser velocity may be generally smaller than the possible jet velocity, due to entrainment of the pre-shock material \citep[e.g.][]{1993ApJ...414..230M, 2016A&A...585A..71M}. If we assume a jet velocity of $\sim$ 200\,$\kms$ \citep[e.g.][]{2018A&ARv..26....3A}, the precession period should be about 1000 years.
According to the fitted $\phi_{0}$ of about 50$\ad$, the P.A. of the jet at the times conducting the ALMA (2019-2021) and VLBA (2015) observations should be less than $\psi$ (70$\ffad$1) and should continue to decrease, i.e. the jet should be located in the north of the mean axis of the jet and moving  farther to the north (see the sketch of the jet/outflow in Fig. \ref{fig:jet-sch}).
This is further discussed in Section \ref{sec:vlba}.

As we can see in Fig. \ref{fig:CO_SiO}, the wiggles of the jet are not very obvious, especially on the blueshifted side of the jet. However, the fitted precession motion is consistent with the distribution and 3D motions of the water masers (see Section 4). Therefore, combining all the results of ALMA and VLBA, we argue that the jet likely harbors a small amplitude precession.
%Alternative explanation, such as the curved jet is because is tracing an extended shell. discussed in \citet{2012ApJ...756..170Q}. However, we found the

\begin{table}
\caption{The parameters for the jet precession.}
\label{tab:jet-precession}
\centering
%\raggedright
\begin{tabular}{ccccc}     % 7 columns
\hline\hline
 $\psi$  & $\alpha$  &  $\lambda$  & $\phi_{0}$  &  $i$  \\
 (1) & (2)  & (3)  &  (4)  &  (5)   \\
 \hline\hline
  70$\ffad$1$\pm$0$\ffad$13 & 0$\ffad$06$\pm$0$\ffad$003  & 7$\ffas$2$\pm$0$\ffas$3  &  50$\ffad$1$\pm$10$\ffad$6 &  8.0 \tablefootmark{$\dag$} \\
 \hline
\end{tabular}
\tablefoot{Column 1: the position angle of the jet symmetry axis in the plane of the sky. Columns 2 and 3: the amplitude and wave scale of the precession. Column 4: the initial phase at the location of MM1. Column 5: the inclination angle of the jet to the plane of the sky.\\
\tablefoottext{$\dag$} {The inclination $i$ is fixed to 8$\ad$ according to the calculations using water masers (see Section \ref{sec:vlba}). This is because the precession fitting is relatively insensitive to the inclination \citep[e.g.][]{2009ApJ...698..184W}.} }
\end{table}

\subsection{Jet kinematics}
\label{sec:jetKin}

\begin{figure*}
    \centering
    \includegraphics[width=0.49\hsize]{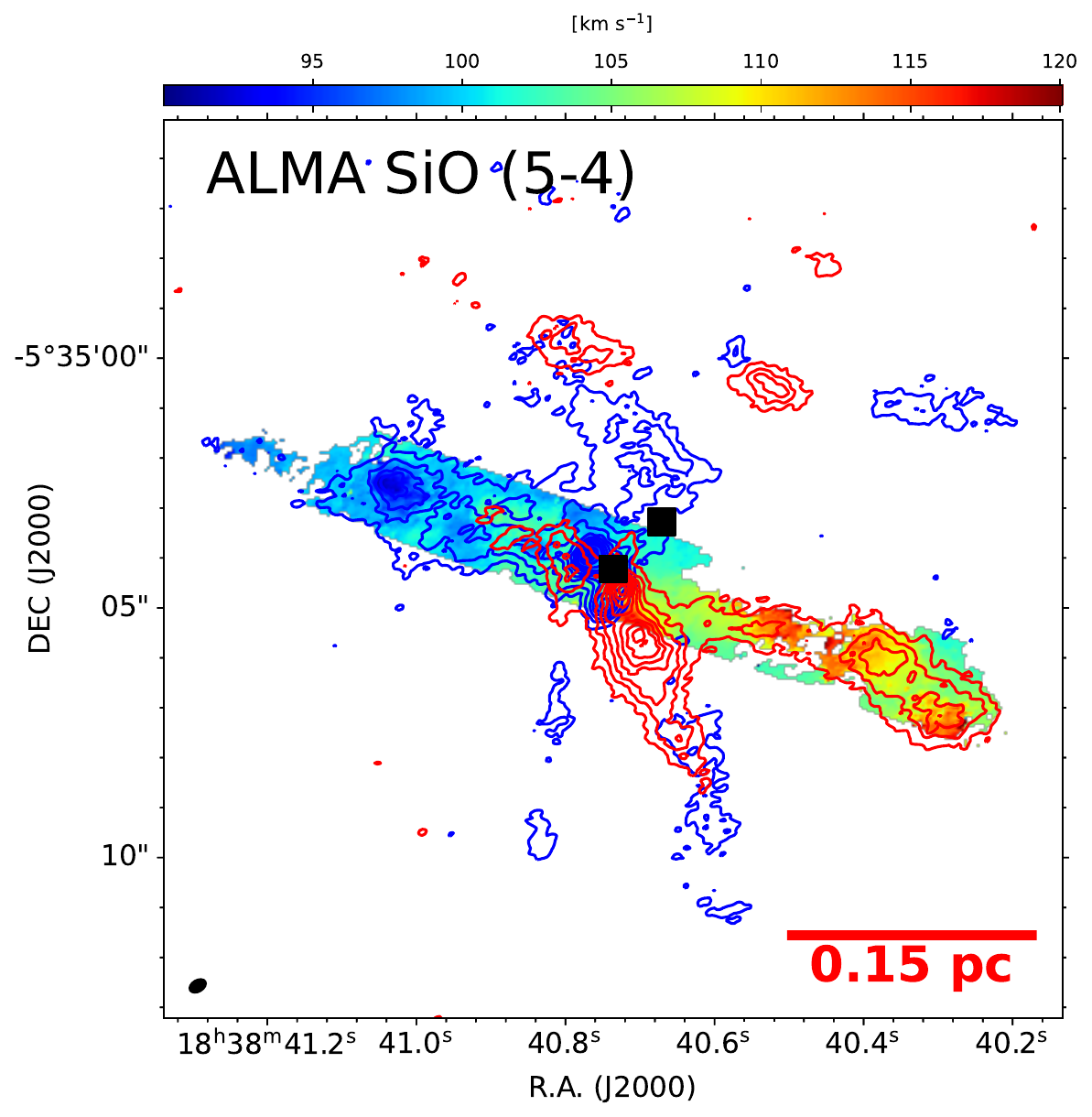}
    \caption{SiO(5-4) intensity-weighted velocity (moment-1, color image) of the jet.
    Only the moment map in the 2$\ffas$0-wide belt along the jet symmetry axis is shown to highlight the kinematics in the collimated jet. The filled squares denote the peak positions of MM1 and MM2. The red and blue contours overlaid are the same as the ones in the right panel of Fig. \ref{fig:CO_SiO}. }
    \label{fig:SiO-mom1}
    \end{figure*}

Jets have been proposed to be the agent to transfer the angular momentum of protostars and disks to outflows. Direct observational evidence has been found in low-mass protostellar jets, e.g. in terms of the systematic velocity gradients perpendicular to the jet axis \citep[e.g.][]{2017NatAs...1E.152L}.
Fig. \ref{fig:SiO-mom1} shows the intensity-weighted velocity (moment-1) SiO (5-4) map of the jet in G26.50+0.28. Only the 2$\ffas$0-wide belt along the jet symmetry axis is shown to highlight the kinematics in the collimated jet.
It seems that the denser clumps in the red/blue jet show more redshifted/blueshifted velocities, which may be caused by an episodic ejection showing jagged velocities \citep[e.g.][]{2001ApJ...551L.171A}.
Besides the knots, no clear velocity trend, i.e. a roughly constant velocity perpendicular to the jet axis is seen in Fig. \ref{fig:SiO-mom1}.

That the jet is collimated and has a small inclination relative to the plane of the sky allows investigations of its transverse kinematics, like in the case of the jet associated with the low-mass protostellar counterpart HH\,212 \citep{2017NatAs...1E.152L}.
Meanwhile, the ALMA observations have moderately resolved  the width and the velocity distribution of the jet.
%\textbf{QUESTION: WHAT do the ALMA data show with respect to the velocity gradient across the jet?}
Therefore, the velocity gradient perpendicular to the axis of the jet, i.e. jet rotation, may not be prominent in G26.50+0.28.

\section{Water maser spots revealed by 15-epoch VLBA data}
\label{sec:vlba}

\subsection{The reference spot}

\begin{figure*}
    \centering
    \includegraphics[width=0.49\hsize]{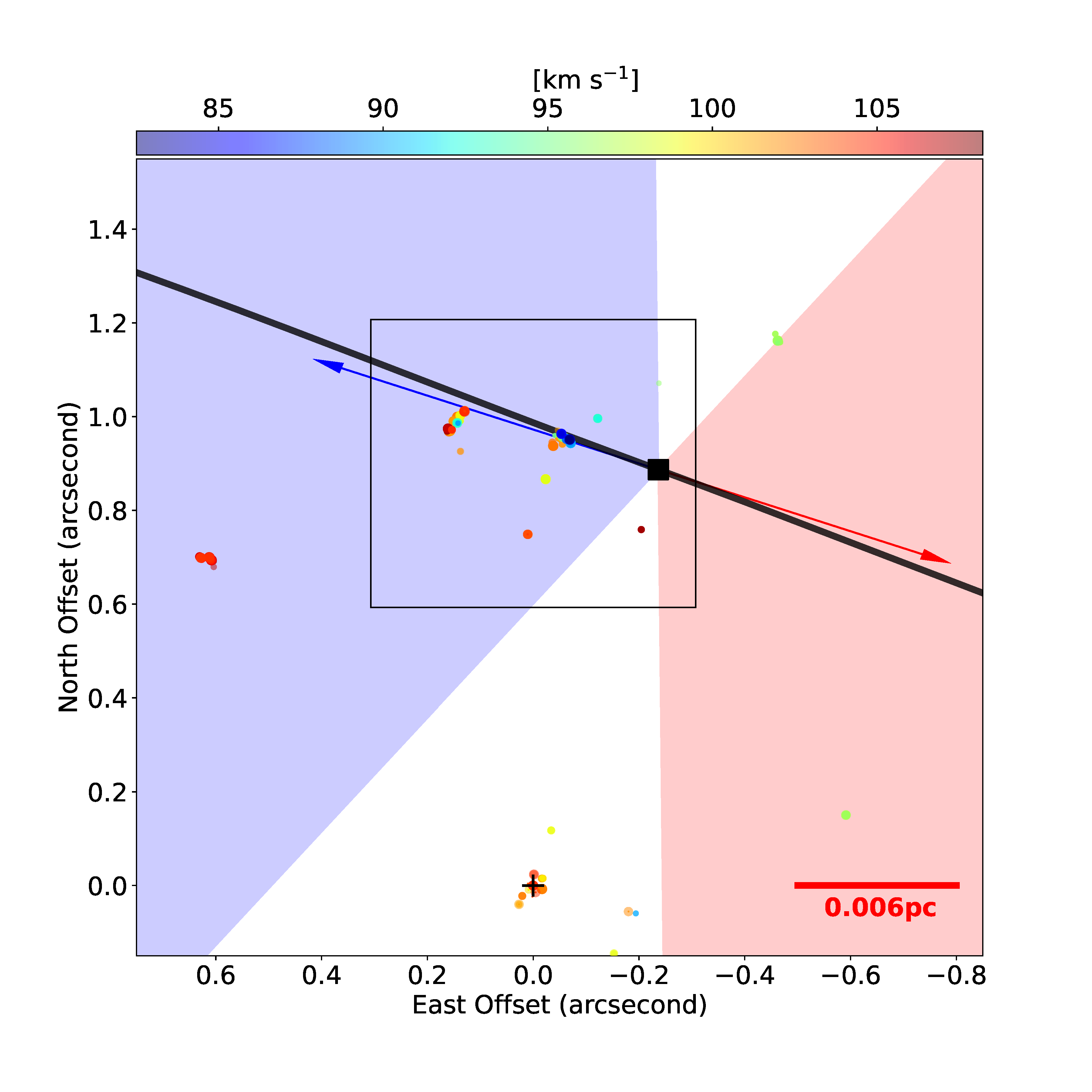}
    \includegraphics[width=0.49\hsize]{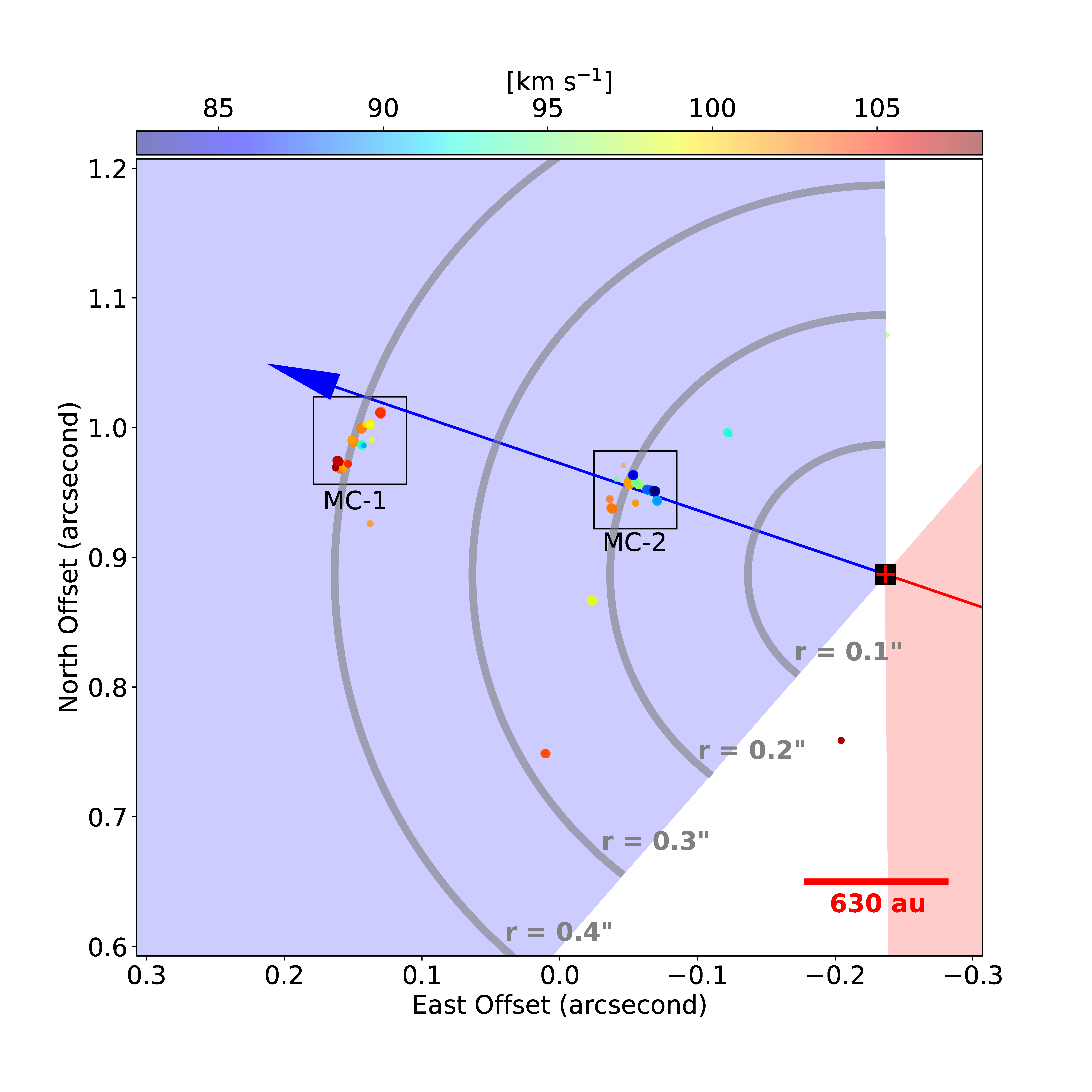}
    \caption{Water maser spots detected by the VLBA observations. Panel (a): The plot of all the maser spots detected with the 15 epoch VLBA observations. Panel (b): The zoom-in image of the black box in panel (a). The reference maser spot at (0, 0) is illustrated by the black cross in the left panel.  In all the panels the colors of the spots represent their line-of-sight velocities and the sizes are proportional to their fluxes. The blue and red arrows in the two panels indicate the jet symmetry axis  of the jet determined by the jet precession fitting (see Section \ref{sec:jet-precession}). The filled square in each panel  denotes the peak positions of MM1. The red cross in the right panel represents the fitted position error of MM1, which is about 0$\ffas$014.
    The blue and red shadowed regions denote the outflow regions, which are the regions outlined by the gray lines in Fig. \ref{fig:CO_SiO}.
    The black line in the left panel shows the loci of the precessing jet (as in Fig. \ref{fig:CO_SiO}).}
    \label{fig:maserDis}
    \end{figure*}

The left panel of Fig. \ref{fig:maserDis} presents the plot of all (3448 in total) the water maser spots  detected in the 15-epoch VLBA data.
As mentioned in Section \ref{obs}, fringe fitting for all the epochs were performed using a strong and point-like water maser spot.
Therefore, all measured proper motions of maser spots are relative to the reference spot.
%%The selected reference spot may have its own proper motion, which leads to overall offsets in the magnitudes and directions of the proper motions for all the maser spots.
The systemic motion can usually be derived by subtracting the mean or model (e.g. fitting the kinematics of the maser spots assuming an isotropic expansion) velocity of the persistent maser spots (found in all epochs) to transfer to a reference frame in which the mean proper motion is zero \citep[e.g.][]{2003ApJ...598L.115T, 2013ApJ...775...79Z}.
It is not applicable in G26.50+0.28 since the maser spots are mainly located in the blue lobe of the outflow, which likely biases the mean motion.

In this specific case, (1) the line-of-sight (l.o.s.) velocities of the reference spot (V$_{lsr}$ = 104.11 $\kms$) and the spots around it are close to the systematic velocity ($\approx$104 $\kms$, derived from Gaussian fits of ALMA CH$_{3}$CN lines \citep[e.g.][]{2012ApJ...756..170Q}); (2) the reference spot is well separated from the main star formation region and the outflow lobes (see the left panel of Fig. \ref{fig:maserDis}).
The reference maser spot may not be (heavily) affected by the star-forming activities and the outflow.
Therefore, the proper motions relative to the spot may trace the internal motions.
The fact that the derived proper motions of the masers are naturally aligned along the direction of the jet identified in the ALMA data  (see Section \ref{sec:pm}) gives our approach an additional measure of robustness.

In brief, for this unique case, we have a co-moving local ``observer'' to monitor the internal velocity of the maser spots, significantly eliminating the systemic motions, e.g., Galactic rotation as well as its non-circular (peculiar) velocity components.

\subsection{The distribution of the water maser spots}

We can see from the left panel of Fig. \ref{fig:maserDis} and Table \ref{tab:msNum} that most (2747, $\approx$ 80\%) of the maser spots are located in the blue lobe of the outflow (the blue shadowed region in Fig. \ref{fig:maserDis}).
Why very few (14) spots are located in the red lobe is not clear. It may be caused by a non-symmetric distribution, since the SiO emission at the base of the redshifted jet is also relatively weak (see the right panel of Fig. \ref{fig:CO_SiO}).
Outside the outflow lobes, maser spots are located mainly around the reference spot (473, $\approx$14\%).
These maser spots may be associated with shocks in the disk, e.g., accretion flows.

In the following, we focus on the maser spots in the blue lobe.
We can see from the right panel of Fig. \ref{fig:maserDis} that the overall distribution of these maser spots presents clearly two patterns, i.e., two concentrated clusters, MC-1 and MC-2, exactly associated with the collimated jet identified in the ALMA data and scattered spots interspersed in the blue lobe.
This two-pattern distribution is consistent with the collimated jet and wide-angle wind scenario seen in the ALMA data.
Along the jet axis, the water maser spots are exclusively distributed in the two clusters, suggesting possible episodic ejection since the masers are likely tracing shocks.

If we naively believe that the flux of maser spots is directly proportional to the intensity of the wind components\footnote{We should note that this assumption is relatively coarse, since the excitation and intensity of masers depends on many factors like the geometry, path lenghts, and the properties of shocks and the local gas \citep[e.g.][]{1992ARA&A..30...75E}.}, the well-distinguished patterns of maser spots enable us to tentatively estimate the relative intensity or the fraction of total accelerated particles between dense (jet) and wide-angle winds for the first time. We can see from Table \ref{tab:epoch} that the total flux of the maser spots associated with the dense jet, i.e. the ones in MC-1 and MC-2, is about the 60\% (median) of the total flux of the maser spots in the blue lobe. That is about 60\% of the accelerated particles (winds) are concentrated (or collimated) to the relative small angle around the jet axis.

\begin{figure*}
    \centering
    \includegraphics[width=0.49\hsize]{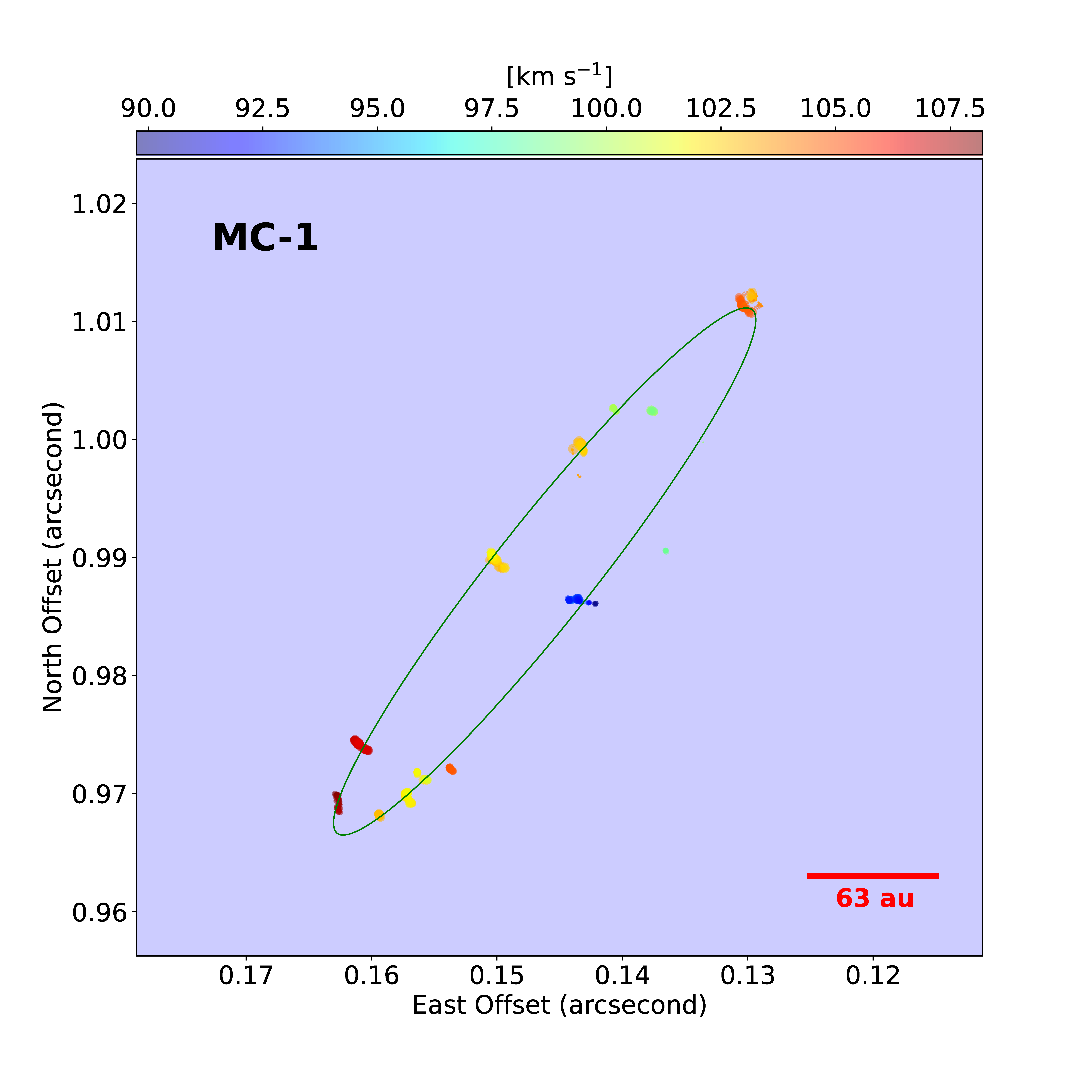}
    \includegraphics[width=0.49\hsize]{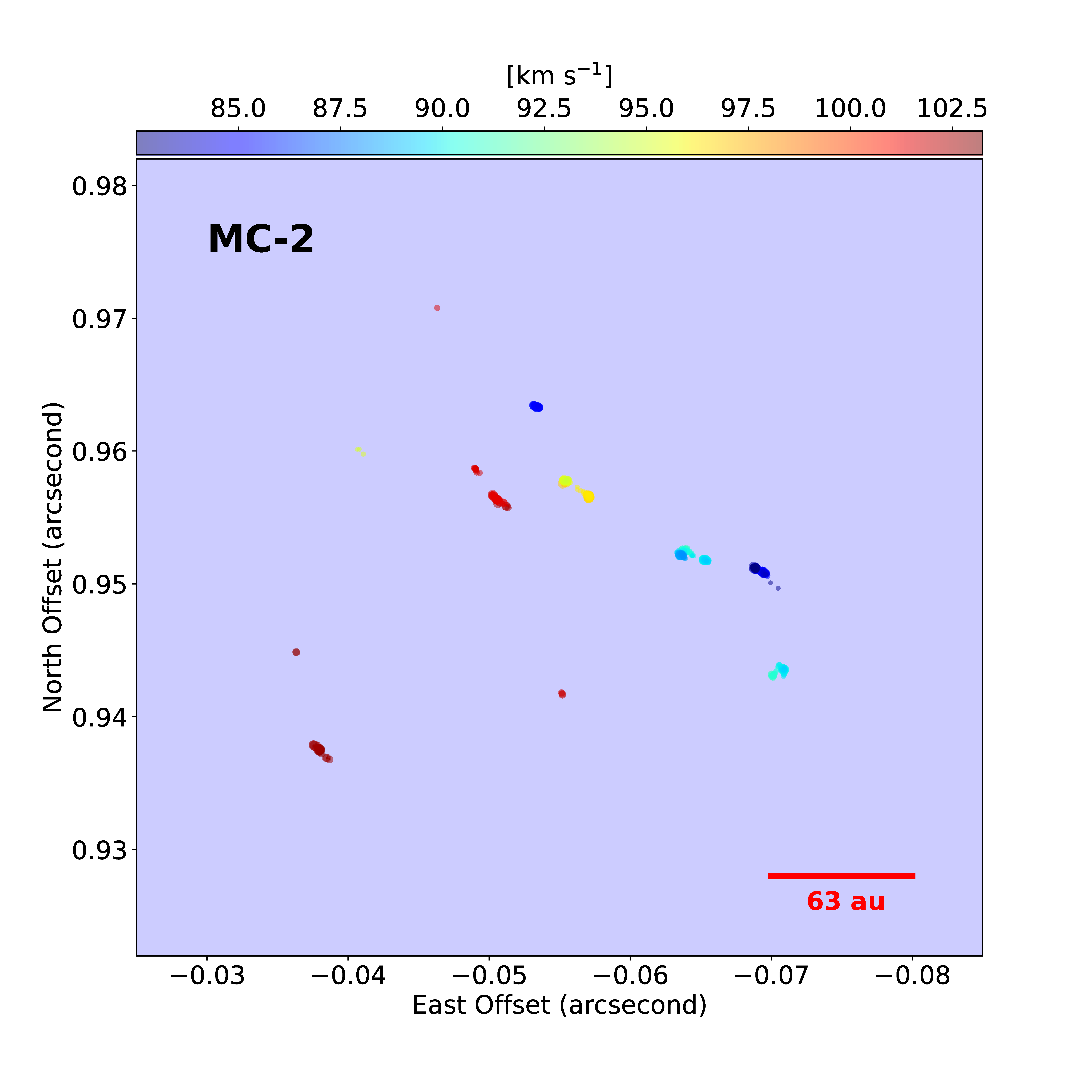}
     \caption{Same as Fig.\ref{fig:maserDis} but zooming into the regions that are illustrated by the black rectangles in the right  panel of Fig.\ref{fig:maserDis}.
     The green ellipse illustrates the fitting result of the ring-like MC-1 maser distribution.}
    \label{fig:mc12}
    \end{figure*}

From the right panel of Fig. \ref{fig:maserDis}, we can see that the P.A. of MC-2 is smaller relative to that of MC-1 (MC-2 is closer to the north direction). This is consistent with the jet precession derived by ALMA data with $\phi_{0}$ of  50$\ffad$1 (see Section \ref{sec:jet-precession}).
However, according to the jet precession fitting in Section \ref{sec:jet-precession}, the two clusters should be all located on the northern side of the symmetry axis (blue arrow in Fig. \ref{fig:maserDis}), which is not true for MC-1 and MC-2.
A possible explanation is that the driving star of the jet may not necessarily be located at the dust peak of MM1 (it then may be located slightly to the south of the peak of MM1).
%
%The maser spots with distances less than 0$\ffas$3 relative to the jet origin (MM1) in the blue lobe seem to be symmetric to the jet axis, which may be simultaneously generated by a burst of the (episodic) winds. The spots in the southern part of the blue lobe seem to be farther away from MM1, but this may also be the consequence of uncertainties in the driving protostellar location.
%
%%%%%%  Andreas: I'm not really sure what you want to say here. There seem to be only 3 spots which are not associated with Mc-2, and they are not really symmetric. Maybe you should remove this part?
%
Fig. \ref{fig:mc12} shows the zoom-in images of MC-1 and MC-2.
It is interesting to see that all the spots of MC-1 are located in a ring-like structure.
We fit these maser spots with a 3D ellipse and the fitting result is shown in the left panel of Fig. \ref{fig:mc12}.
The fitted major and minor, $a$ and $b$, axes are 27$\ffas$6 and 4$\ffas$0, respectively.
Assuming the ring is a 3D circle and perpendicular to the jet axis, we can directly derive the inclination of the jet, $i = \arcsin(b/a)$ $\approx$ 8$\ffad$4$\pm$0$\ffad$5.
Meanwhile, we also fit the spots of the first seven (E2--E8) and last seven epochs (E10--E16) with 3D ellipses and do not find clear expanding or contracting motion of the ring within the time range of about one year (see Table \ref{tab:epoch}).

\begin{figure*}
    \centering
    \includegraphics[width=0.49\hsize]{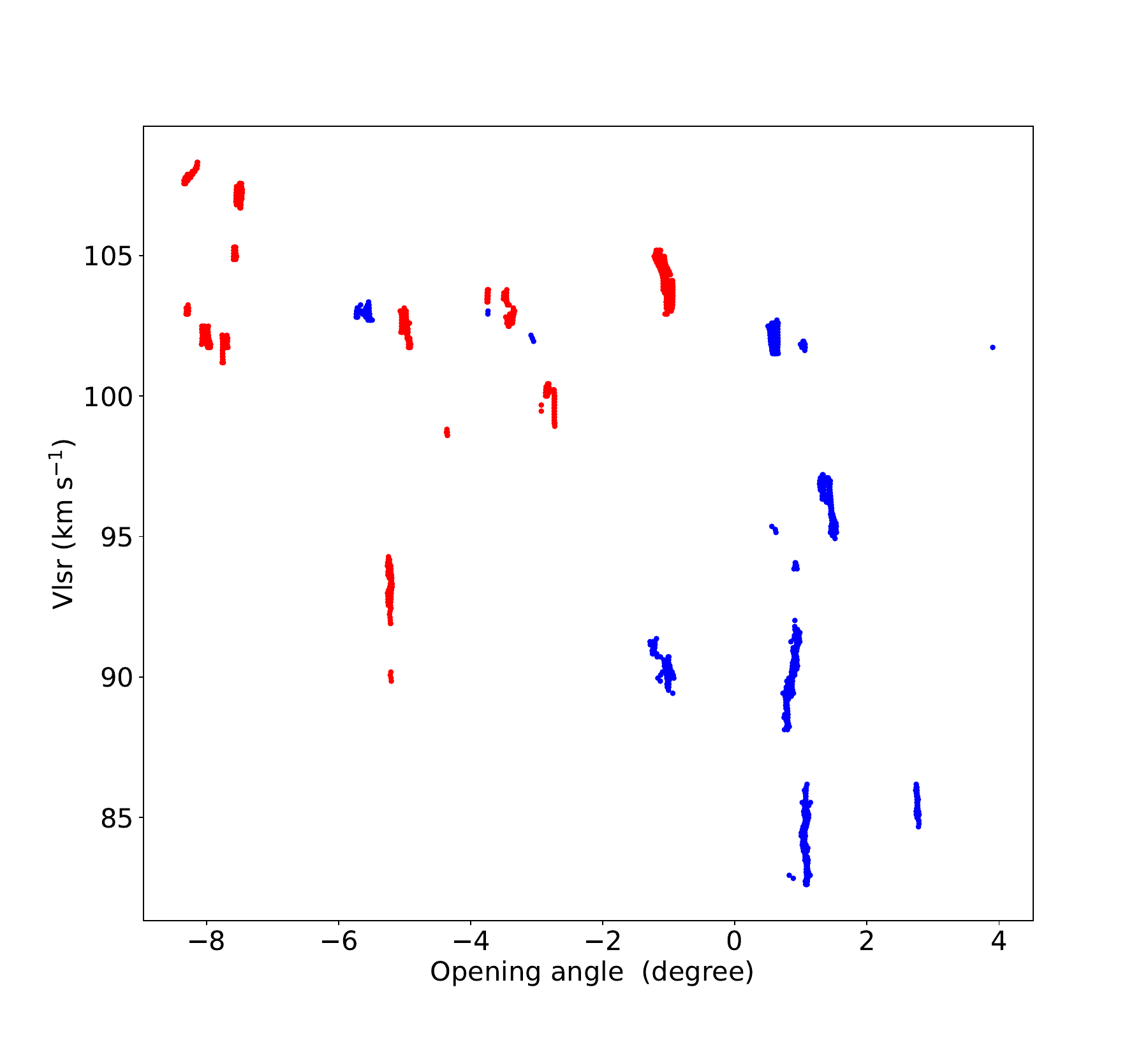}
    \includegraphics[width=0.49\hsize]{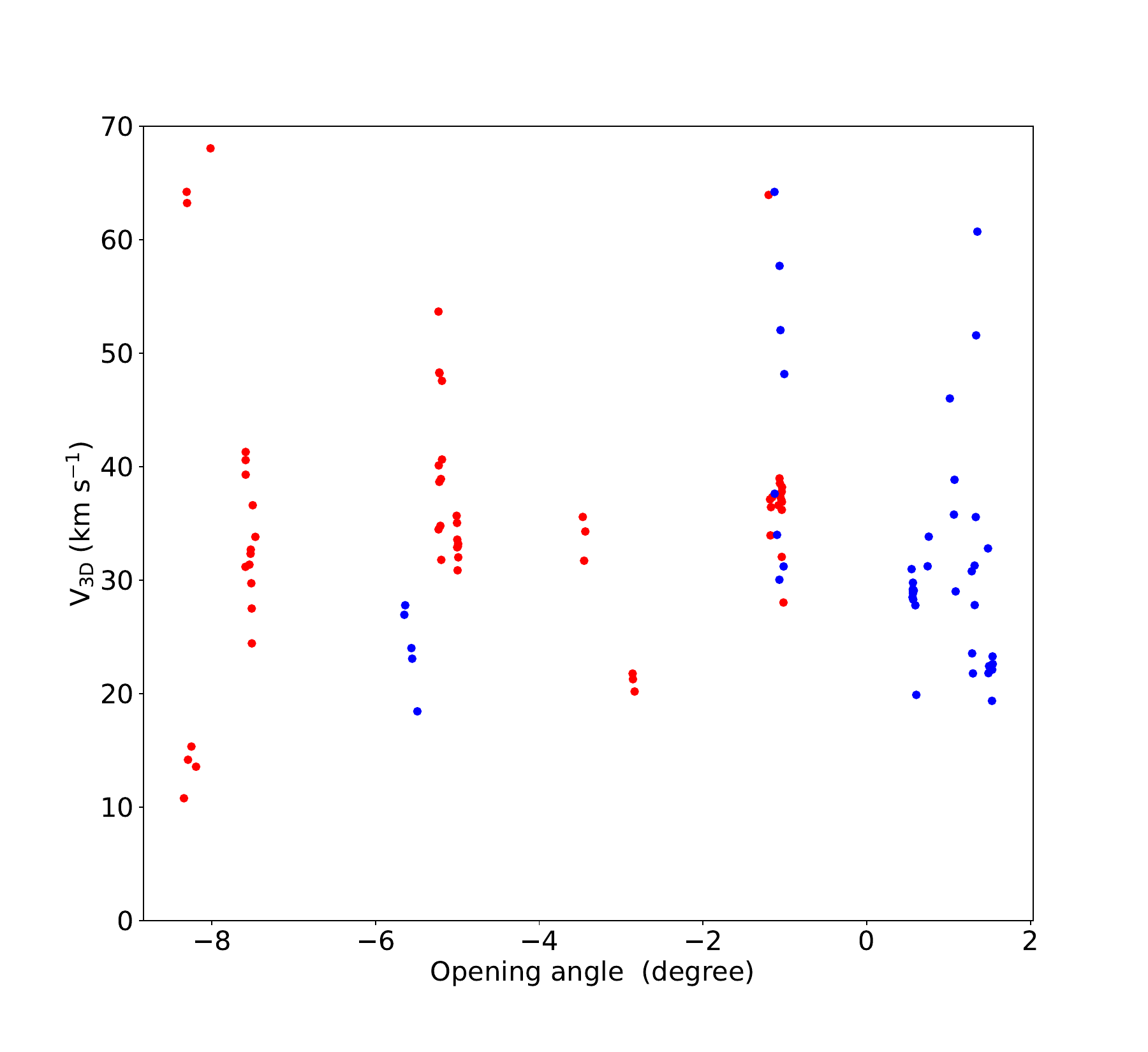}
    \caption{Line-of-sight (the left panel, $V_{\rm lsr}$)  and  3D jet velocities (the right panel, $V_{\rm 3D}$) of the maser spots in MC-1 (red dots) and MC-2 (blue dots) against their opening angles relative to the jet axis. Angles measured in counter-clockwise and clockwise directions are negative and positive, respectively.}
    \label{fig:mc12-vlsr}
    \end{figure*}

As mentioned in Section \ref{sec:jetKin}, no clear velocity gradient, i.e. jet rotation, has been found in the SiO moment-1 map (Fig. \ref{fig:SiO-mom1}), which is also true for the l.o.s. velocities (the color of the spots) of the maser spots in MC-1 and MC-2 (see Fig. \ref{fig:mc12}).
The left panel of Fig. \ref{fig:mc12-vlsr} shows the l.o.s. velocity of the maser spots in MC-1 (red) and MC-2 (blue) against the opening angle of the spot relative to the jet axis.
Due to the large dispersion, no clear velocity trend is found in either MC-1 or MC-2.
The overall velocity difference between the two clusters is likely caused by the precession.
The median velocities along l.o.s. of MC-1 and MC-2 are 102.2 and 93.5 $\kms$. Complemented by $\phi_{0}$ of 50$\ad$ (see Table \ref{tab:jet-precession}), the jet precession is counterclockwise when looking along the blueshifted jet axis from MM1, which is co-rotating with the disk/toroid.

\begin{table}
\caption{Numbers of  water maser spots in different regions.}
\label{tab:msNum}
\centering
\begin{tabular}{cc}     % 7 columns
\hline\hline
 Region \tablefoottext{$\dag$}   & Number   \\
 (1) & (2)     \\
 \hline\hline
Total     &  3448  \\
blue lobe &  2747  \\
red lobe  &  14  \\
MC1       &  871  \\
MC2       &  816  \\
Ref\tablefoottext{$\dag$}       &  473  \\
\hline
\end{tabular}
\tablefoot{
\tablefoottext{$\dag$}{The maser spots in the blue and red lobes represent the ones in the blue and red shadowed regions in Fig. \ref{fig:maserDis}.
The spots in MC-1 and MC-2 represent the ones in the two panels of Fig. \ref{fig:mc12}.
Spots in the Ref region are the ones around ($\la$0$\ffas$1) the reference spot identified at V$_{lsr}$ = 104.11 $\kms$ (channel 1000) and (R.A., DEC)$_{\rm J2000}$ = (18:38:40.1707, -05:35:42.7749).}
}
\end{table}

\subsection{Proper motions}
\label{sec:pm}

\begin{figure*}
    \centering
    \includegraphics[width=0.49\hsize]{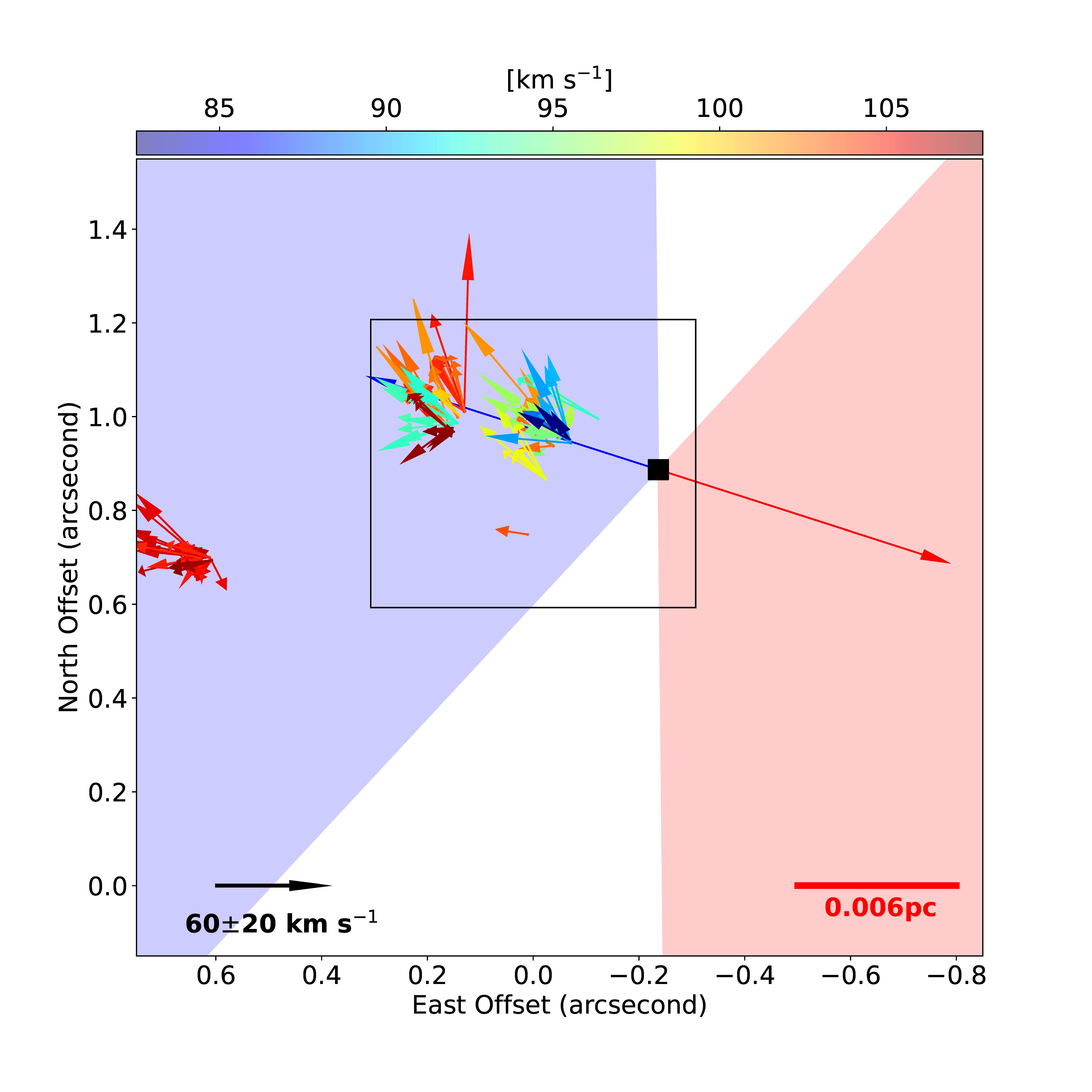}
    \includegraphics[width=0.49\hsize]{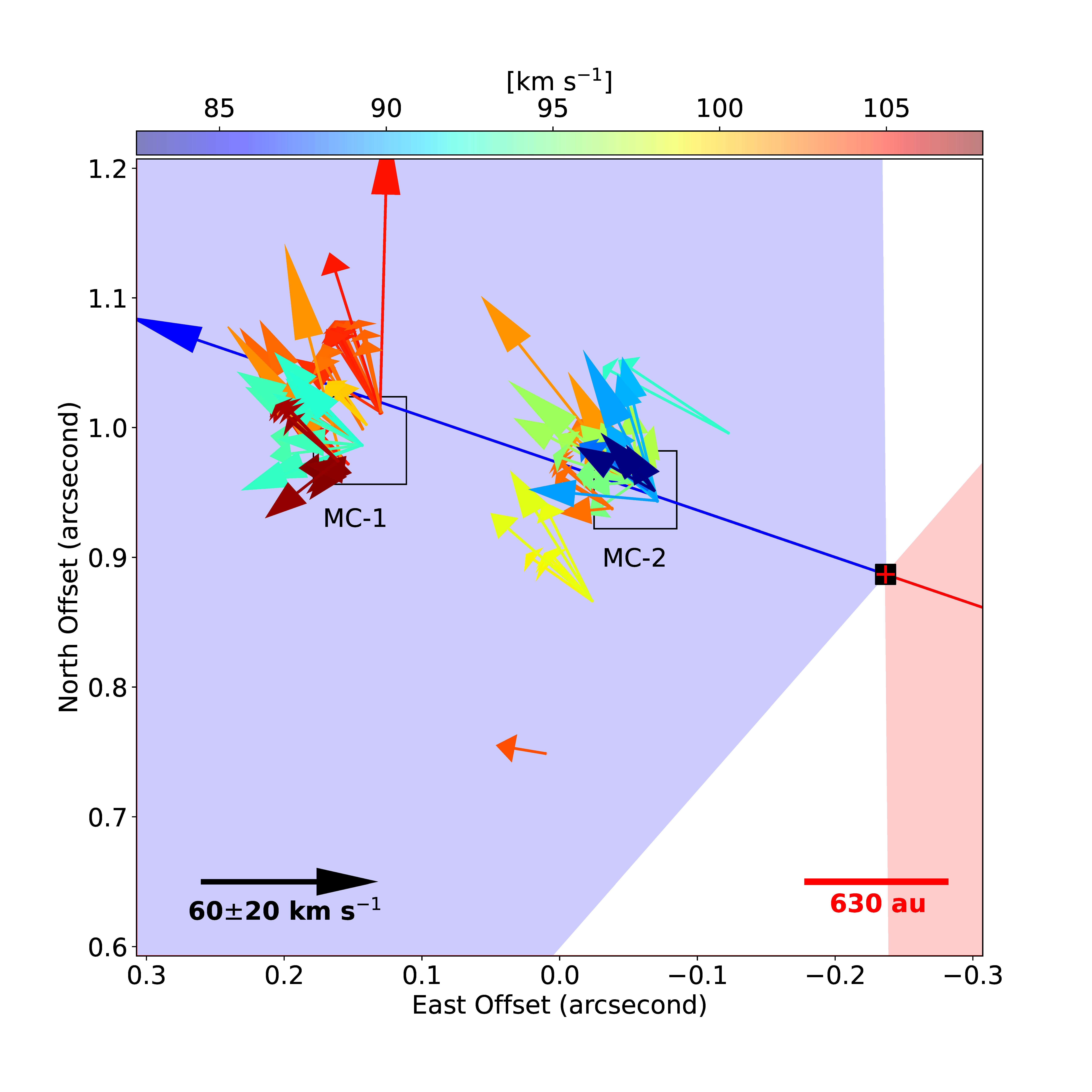}
    \includegraphics[width=0.49\hsize]{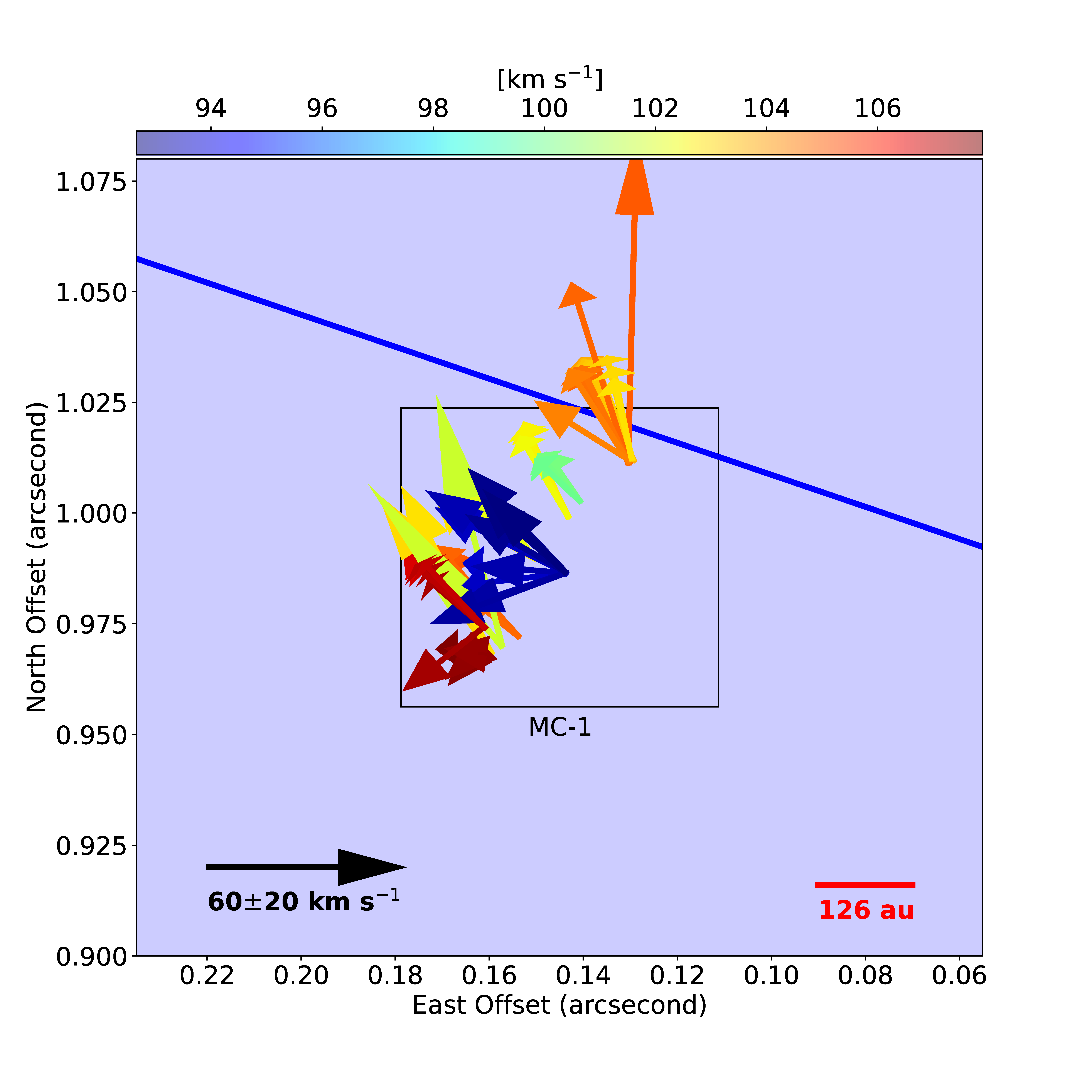}
    \includegraphics[width=0.49\hsize]{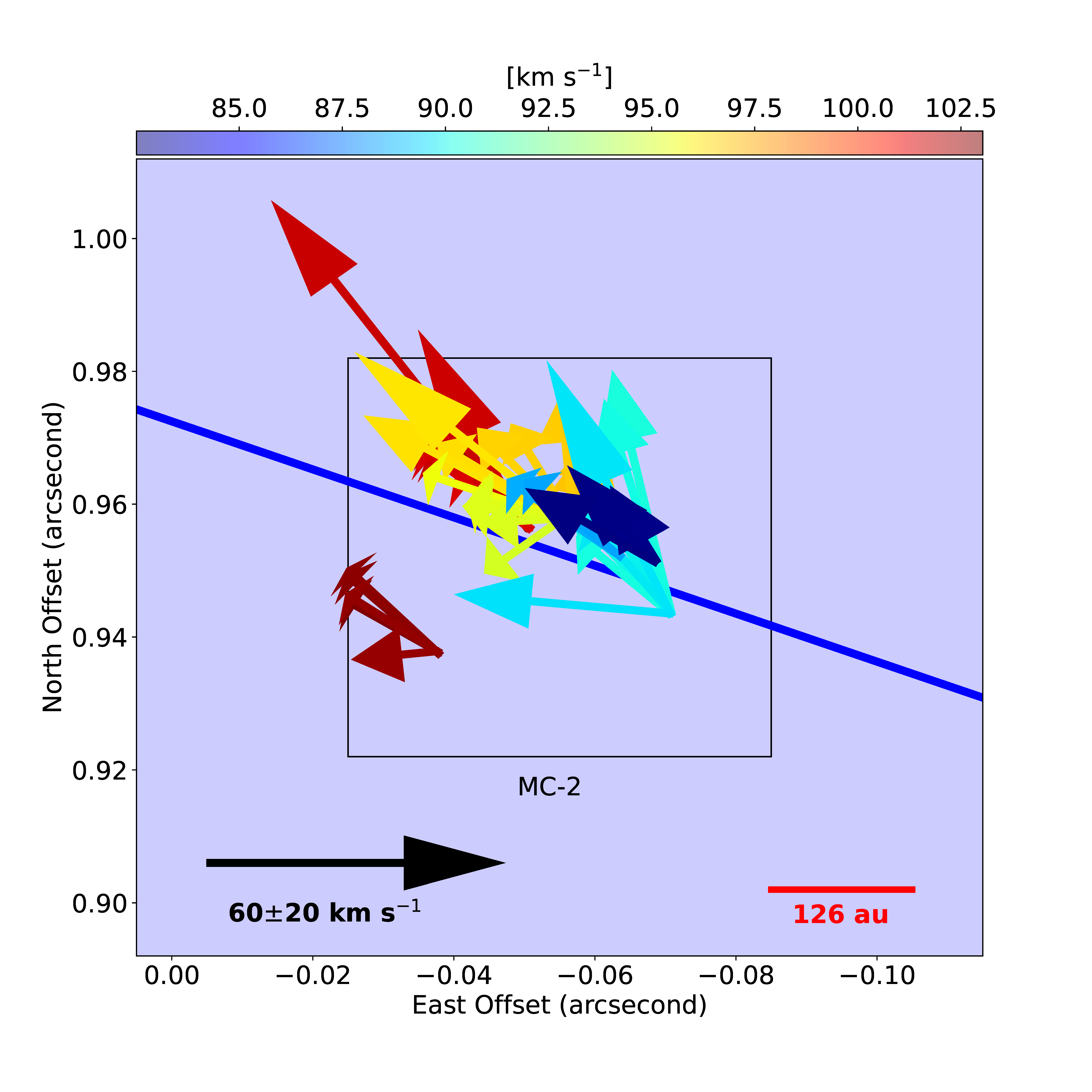}
    \caption{Proper motions derived with 15-epoch VLBA observations. Left panel: Plot of the proper motions of spots persisting more than five epochs. The arrows demonstrate the proper motions of the maser spots. The length (including head), direction, and the head length of the arrow illustrate the magnitude, direction, and error of the proper motion.   Right panel: Same as the left panel but zooming into the region that is illustrated by the black rectangle in the left panel. The filled square in each panel denotes the peak positions of MM1. The red cross in the top right panel represents the fitted position error of MM1, which is about 0$\ffas$014. Black rectangles illustrate the zoom-in regions of Fig. \ref{fig:mc12}.
    Bottom panels: Same as the top two panels but for the two maser clusters MC-1 and MC-2. }
    \label{fig:pm}
    \end{figure*}

\begin{figure*}
    \centering
    \includegraphics[width=0.33\hsize]{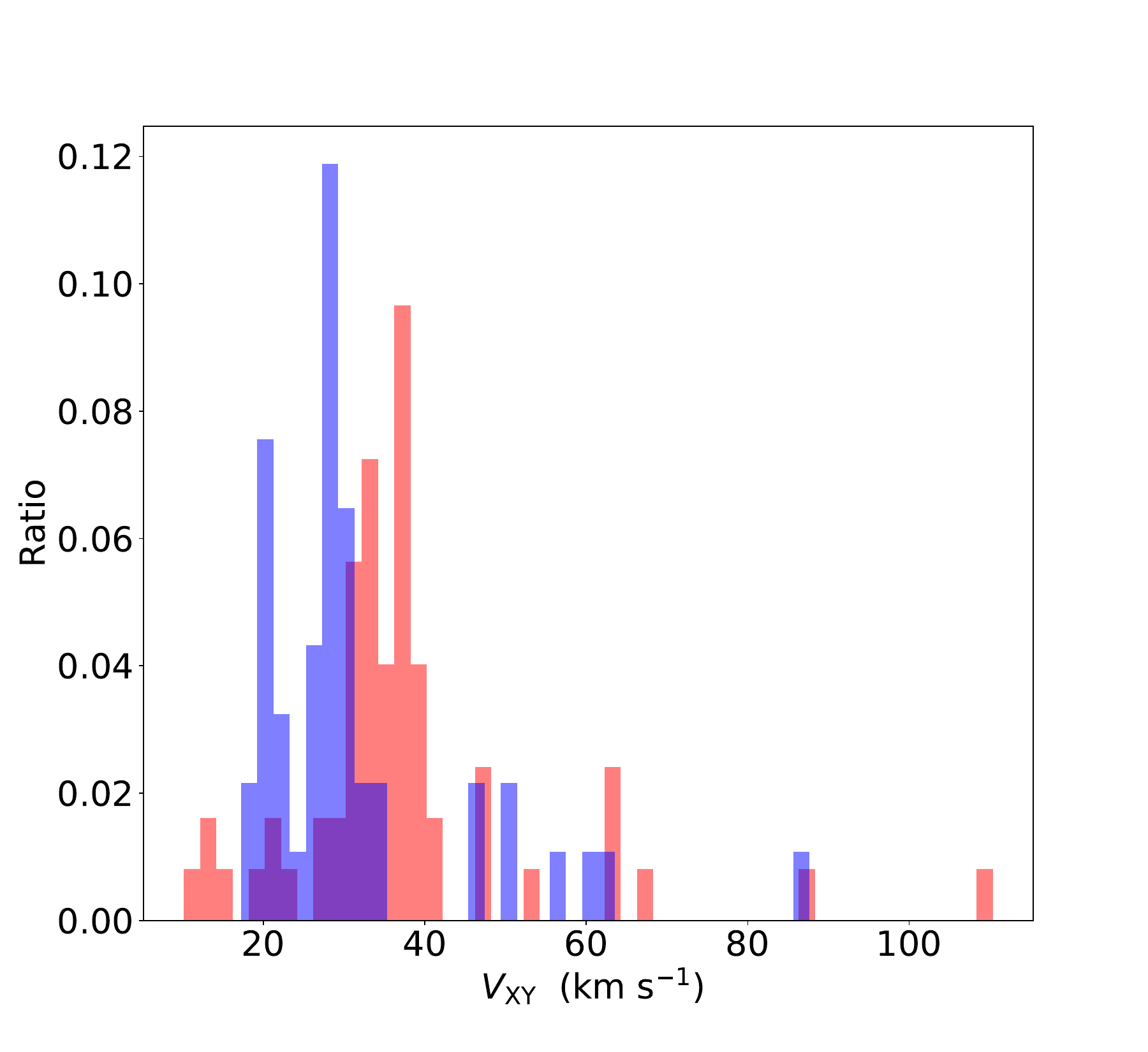}
    \includegraphics[width=0.33\hsize]{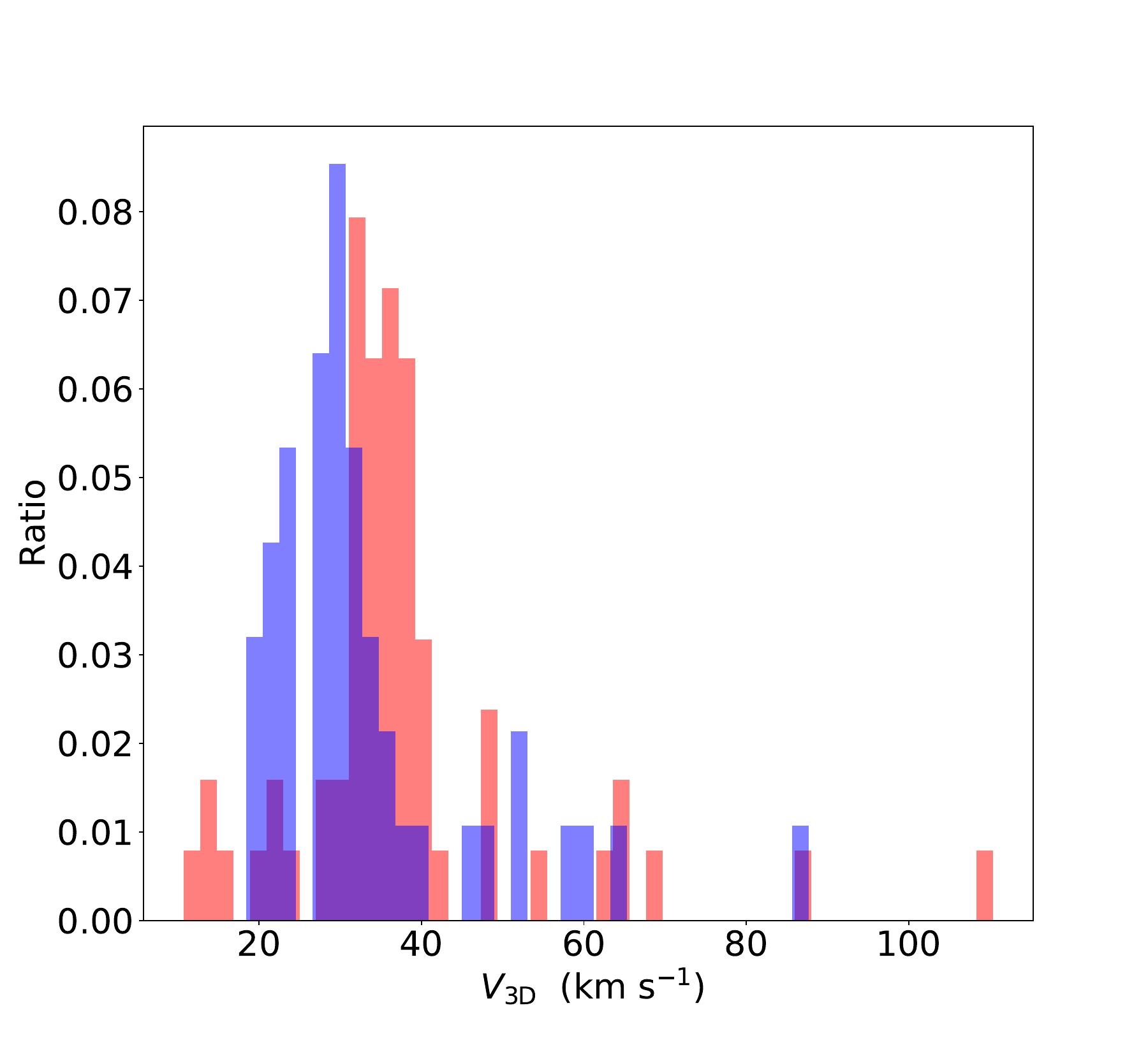}
    \includegraphics[width=0.33\hsize]{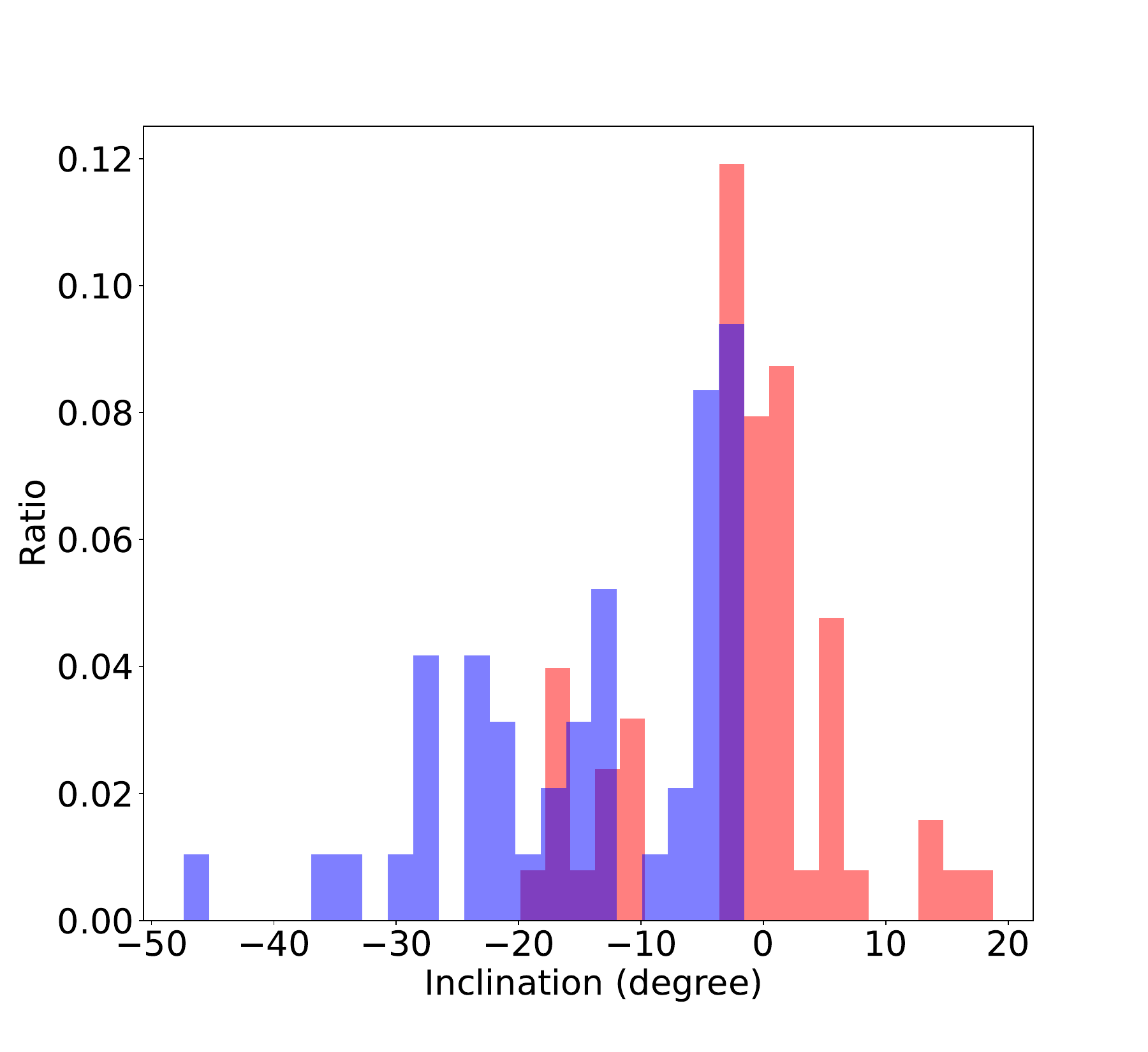}
     \caption{Statistics of the proper motion ($V_{\rm XY}$, left panel), the 3D jet velocity ($V_{\rm 3D}$, middle panel), and the inclination  of MC-1 (red) and MC-2 (blue) with respect to the plane of the sky. }
    \label{fig:histPM}
    % The data comes from Table \ref{tab:table}.\label{fig:general}}
    \end{figure*}

We select 216 spots persisting for at least 5 epochs to calculate their proper motions. A constant velocity (linear) model is used in the fitting, since no clear evidence of acceleration is found.
Fig. \ref{fig:pm} presents the derived proper motions of the spots.
We can see that the proper motions of the spots in the blue lobe of the outflow are exactly consistent with the direction of the jet, especially the spots in MC-1 and MC-2.
From Fig. \ref{fig:histPM} we can see that the overall proper motion ($V_{\rm XY}$) and the 3D velocity ($V_{\rm 3D}$) of MC-1 are all larger than those of MC-2 (see also Table \ref{tab:mc12_pm}).
The 3D velocities are generally larger than those of the high-mass sample in the Protostellar Outflows at the EarliesT Stages (POETS) survey \citep{2020A&A...635A.118M},
suggesting that the O-type protostar may launch winds with even greater intensity.
The 3D velocities also allow us to calculate the inclination of each spot (see the right panel of Fig.\ref{fig:histPM} and also  Table \ref{tab:mc12_pm}).
The median inclinations of the two clusters are 1$\ffad$1$\pm$8$\ffad$2 and 13$\ffad$8$\pm$11$\ffad$0, which are roughly consistent with the result from the elliptical fitting of MC-1 ($\approx$8$\ad$).
Given all the calculations, the typical inclination of the collimated jet may be around 8$\ad$ with respect to the plane of the sky.

Observational evidence to distinguish the X- and disk-wind models is the variation of the magnitude of the velocity with different opening angles \citep[e.g.][]{2011A&A...532A..59A}.
However, studies based on spectroscopic observations usually only reveal velocities along the l.o.s.
%Meanwhile, the result may also be exclusive since velocities are commonly accumulated with the emission of foreground and background gas.
Meanwhile, the line emission of the jet is usually contaminated by the emission of the ambient gas.
On the contrary, the maser spots are from very tiny regions, which should largely eliminate the emission from the background or foreground emission.
Furthermore, 3D motion can all be derived in multi-epoch studies.

We plot the 3D jet velocity of the maser spots in MC-1 and MC-2 against the opening angle in the right panel of Fig. \ref{fig:mc12-vlsr}.
It seems to present a constant distribution of the 3D jet velocity ($V_{\rm 3D}$)  against the opening angle with respect to the symmetry axis of the jet albeit with large dispersion.
The constant distribution may suggest that the winds associated with the collimated jet may be launched from a relatively narrow region of the disk, since the launching velocity is proportional to the Keplerian velocity of the launching point \citep[e.g.][]{2016ARA&A..54..491B}.

At last, we should note that proper motions of water masers may harbor large dispersions \citep[e.g.][]{2006ApJ...645..337H, 2016A&A...585A..71M}. For G26.50+0.28, the distribution and the proper motions of the maser spots, especially in MC-1 and MC-2, are consistent with the collimated jet. Therefore, it is likely that the proper motions of these spots are representing the kinematics of the jet. Meanwhile, the relatively robust medians of the velocities are commonly used in this work, alleviating the possible turbulent motions of the maser spots.

\begin{table*}
\caption{Statistics of the spots in MC-1 and MC-2.}
\label{tab:mc12_pm}
\centering
%\raggedright
\begin{tabular}{ccccc}     % 7 columns
\hline\hline
 Region    &    $V_{\rm 3D}$  ($\kms$)   &    $V_{\rm XY}$ ($\kms$)    &    $V_{\rm Z}$ ($\kms$)  &   inclination  \\
 \hline\hline
 MC-1  &    35.6$\pm$15.7   &   35.5$\pm$15.8     &  -1.02$\pm$4.6  &   1$\ffad$1$\pm$8$\ffad$2    \\
 MC-2  &    29.1$\pm$13.4   &   28.7$\pm$13.6     &  -7.62$\pm$5.9  &   13$\ffad$8$\pm$11$\ffad$0   \\
 \hline
\end{tabular}
\tablefoot{The columns all show the medians of different parameters.  Columns 2, 3, and 4: 3D jet,  plane of the sky (proper motion), and line-of-sight velocities of the maser spots in MC-1 and MC-2. Column 5: Inclinations calculated from the three-dimensional velocity components. }

\end{table*}

\section{Discussion and summary}

We have studied the detailed morphology and kinematics of the outflow in the high-mass star formation region G26.50+0.28.
The ALMA and VLBA observations reveal a clear and consistent picture of the outflow associated with an O-type protostar candidate (see Fig. \ref{fig:jet-sch}), which is a scaled-up version of the collimated jet and wide-angle wind scenario of the low-mass counterpart, e.g. HH\,212 \citep{2017NatAs...1E.152L,2022ApJ...927L..27L}.
The collimated jet is first identified with the ALMA $^{13}$CO (2-1), SiO (5-4), and VLBA water maser spots, which is exactly perpendicular to the elongated disk/toroid and the velocity gradient of the disk.
The jet is embraced by cavity shells with an opening angle of about 140$\ad$. The enhanced north-south shell is likely caused by the interaction of the wind with the ambient filamentary structure.

Regarding the directions of the outflow as defined by \citet{2012ApJ...756..170Q} and this work (see Section \ref{sec:CO_SiO}), there is a $\approx$25$\ad$ disparity between the directions of our collimated jet (also the mean outflow direction) and the axis of the bipolar outflow reported by \citet{2012ApJ...756..170Q}.
This introduces a caveat to the studies based on directions of jets/outflows \citep[e.g.][]{2014ApJ...792..116Z, 2017ApJ...846...16S}, especially in high-mass star-forming regions with extremely wide outflows, where the basic results  may depend on the chosen resolution and tracer.

The collimated jet is likely suffering a small amplitude counterclockwise  precession, when looking along the blueshifted jet axis from MM1, with a precession length of 0.22pc.
Assuming a 30 $\kms$ jet velocity, the possible period is on the order of 7000 years.
%
%\textbf{As previously mentioned, the outflow velocity of the water masers may be lower than the actual jet velocity due to the entrainment of pre-shock material. \citep[e.g.][]{1993ApJ...414..230M, 2016A&A...585A..71M}. The 3D velocity of the two water maser clusters,$V_{\rm 3D}$, is $\ga$ 30\,$\kms$ (see Table \ref{tab:mc12_pm}), indicating that the precession period should be less than about 7000 years.}
%
%
The inclination of the collimated jet should be very low ($\approx$8$\ad$), which makes the jet in G26.50+0.28 a promising target to study the transverse morphology and kinematics of the jet and also the interactions with the edge-on disk \citep[e.g.][]{2017NatAs...1E.152L, 2019Natur.565..206S}.
However, no clear evidence of a velocity gradient perpendicular to the jet axis, i.e. jet rotation along the axis, is found in ALMA or VLBA data.

Water masers are mainly located in the blue lobe of the outflow. Two water maser clusters are closely associated with the collimated jet.
Thus it is likely they are produced by the dense jet.
The median distances of the two water maser clusters to MM1 in the plane of the sky  are 0$\ffas$39 and 0$\ffas$19, respectively.
Assuming that MC-1 and MC-2 are ejected near the MM1 region and their velocities correspond to the median V$_{XY}$ (see Table \ref{tab:mc12_pm}), the estimated dynamical timescales for MC-1 and MC-2 would be approximately 300 and 200 years, respectively. This suggests that they appear to arise from separate ejections or shocks.
However, their origin, especially the very regularly distributed ring-like MC-1, is relatively elusive.
Higher resolution SiO and radio continuum data would help us to understand their origin.
For example, they could be produced by episodic ejections traversing through a small cloud.
Another possibility is that the two clusters may be generated by different shocks, i.e. the leading and reverse shocks within an internal working surface,
since the leading MC-1 with velocity greater than 30 $\kms$ ($\approx$35.6 $\kms$) may be associated with J-shocks \citep[e.g.][]{1993ApJ...414..230M, 2020A&A...635A.118M} and also moves faster. On the contrary, MC-2 with velocities $\approx$28.8 $\kms$ is slower and may be associated with C-shocks.
%

%\textbf{Assuming that MC-1 and MC-2 are ejected near the MM1 region (their median distances to MM1 of 0$\ffas$39 and 0$\ffas$19) and their velocities correspond to the median V$_{XY}$ (see Table \ref{tab:mc12_pm}), the estimated dynamical timescales for MC-1 and MC-2 would be approximately 300 and 200 years, respectively. This suggests that they are unlikely to share a common origin, but rather represent two separate ejections or different shocks as mentioned earlier.}

%
We also tentatively estimate the amount of accelerated particles by comparing the total maser flux between the dense and diffuse wind components, i.e. the collimated jet and wide-angle winds.
About 60\% of the accelerated particles may be concentrated (or collimated) along the jet with a small opening angle around the jet axis, which means the collimated jet and wide-angle winds are roughly comparable in G26.50+0.28.
The 3D  jet velocities derived from the water maser spots with opening angles $\la$8$\ad$ seem to be constant, suggesting the jet may be launched in a relatively  small region, which favors the X-wind model \citep[e.g.][]{2007prpl.conf..261S}.

\begin{acknowledgements}
This work was funded by the National Key R\&D Program of China (No. 2022YFA1603103), the CAS ``Light of West China'' Program (No. 2021-XBQNXZ-028), the National Natural Science foundation of China (Nos.12103082 and 11603063), and the Natural Science Foundation of Xinjiang Uygur Autonomous Region (No. 2022D01A362).
GW acknowledges the support from Youth Innovation Promotion Association CAS.

\end{acknowledgements}


\begin{appendix}

\onecolumn
\section{Distribution and proper motions of the water maser features}
\label{sec:feature}

   \begin{figure*}[h]
    \centering
    \includegraphics[width=0.45\hsize]{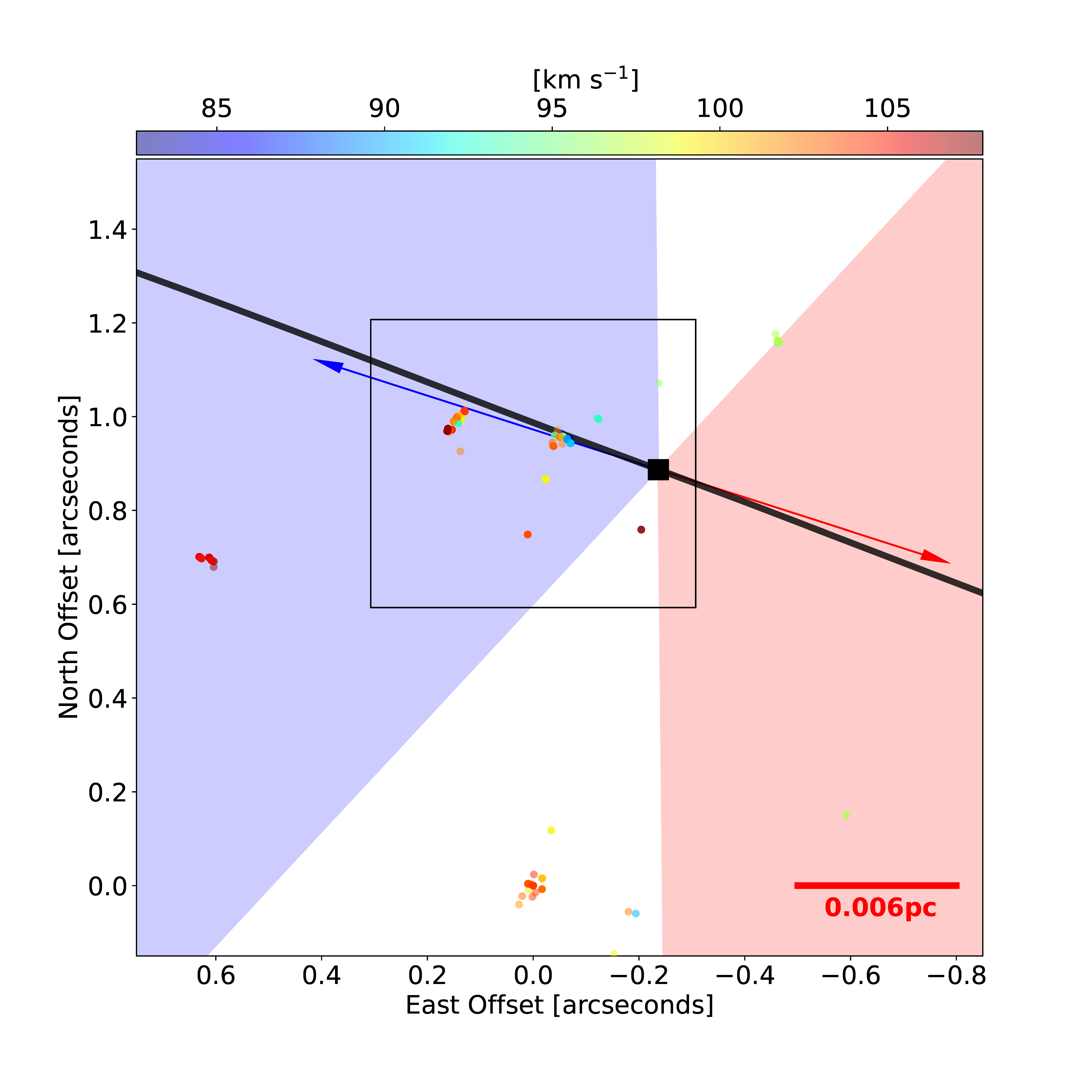}
    \includegraphics[width=0.45\hsize]{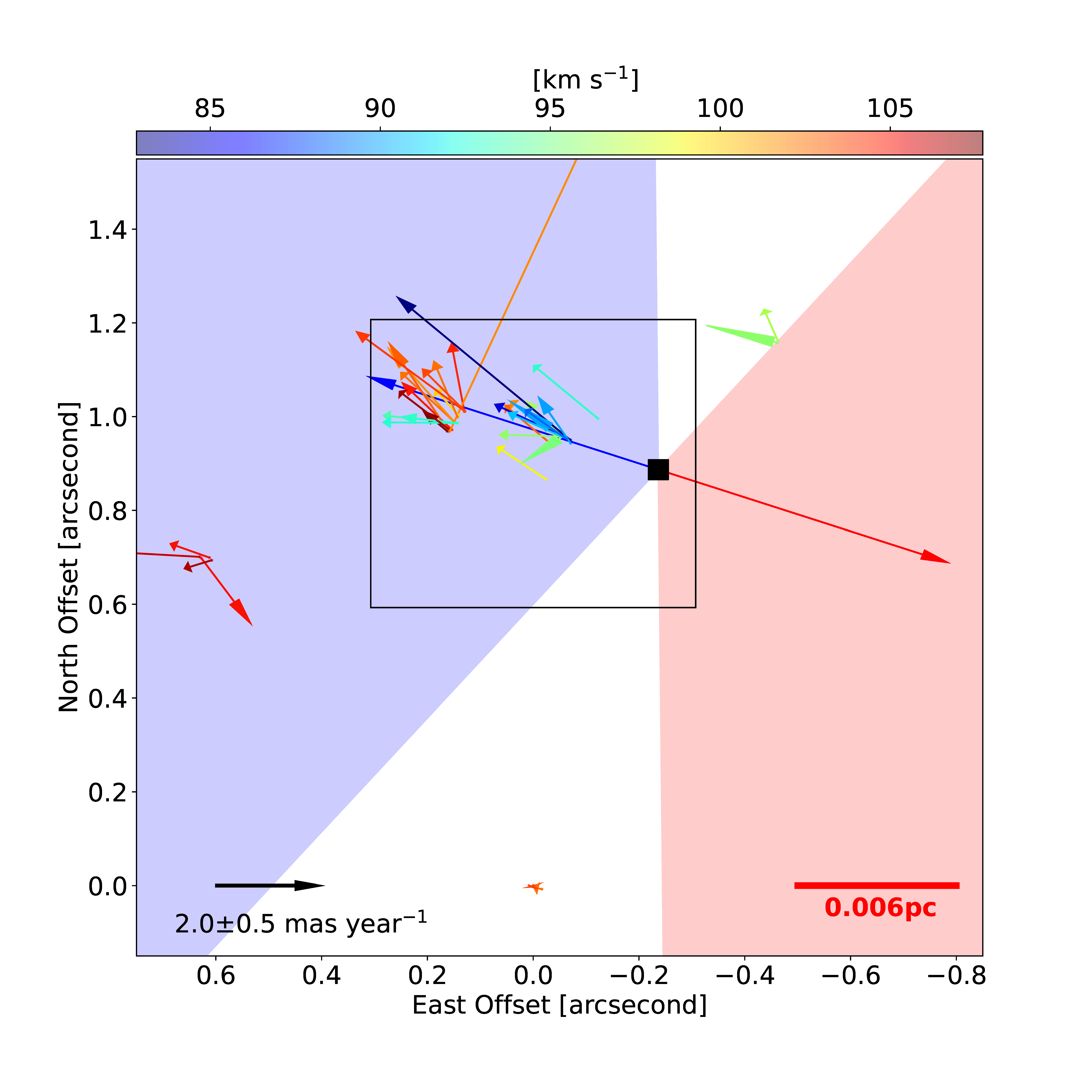}
    \caption{Distribution and proper motions of the water maser features detected by the VLBA observations. Left panel: The plot of all the maser features detected with the 15 epoch VLBA observations. The colors of the dots represent their line-of-sight velocities. The blue and red arrows in the two panels indicate the jet symmetry axis  of the jet determined by the jet precession fitting (see Section \ref{sec:jet-precession}). The blue and red shadowed regions denote the outflow regions, which are the regions outlined by the gray lines in Fig. \ref{fig:CO_SiO}.
    The black line in the left panel shows the loci of the precessing jet already introduced in Fig. \ref{fig:CO_SiO}.
    Right panel: The plot of the proper motions of maser features persisting more than 3 epochs. The arrows indicate the proper motions of the maser features. The length (including head), direction, and the head length of the arrow illustrate the magnitude, direction, and error of the proper motion.
    Black rectangles in the two panels illustrate the zoom-in region of the right panel of Fig. \ref{fig:maserDis}. }
    \label{fig:feature}
    \end{figure*}

\section{ALMA 1.3\,mm continuum image restored with a 0$\ffas$3 circular beam}

   \begin{figure*}[h]
    \centering
    \includegraphics[width=0.45\hsize]{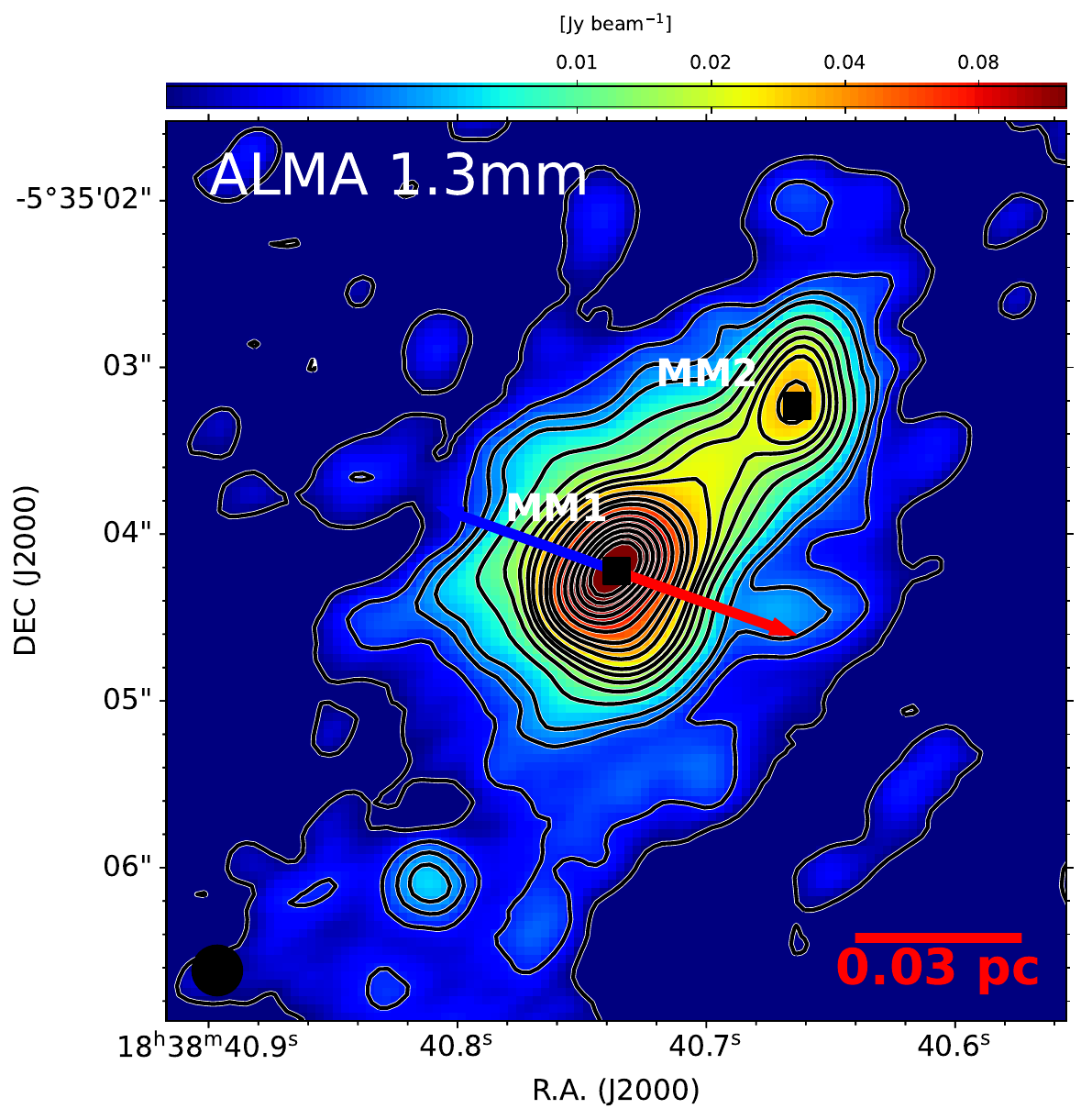}
    \caption{
    ALMA 1.3mm continuum restored with a circular beam of 0$\ffas$3. The contour levels are set to [-3, 3, 6, 9, 12, 20, 30, 40, 50, 60, 80, 100, 120, 160, 200, ..., 360] $\times$ 0.4\,m$\Jypb$. Dashed contour lines barely seen, because having almost disappeared because of smoothing, represent negative values. A filled ellipse in the lower left shows the beams. The blue and red arrows indicate the directions of the jet. The filled squares, denote the peak positions of the two cores which are labeled as MM1 and MM2.}
    \label{fig:cont_bmc}
    \end{figure*}

\section{Low-velocity $^{13}$CO emission of the outflow}

   \begin{figure*}[h]
    \centering
    \includegraphics[width=0.45\hsize]{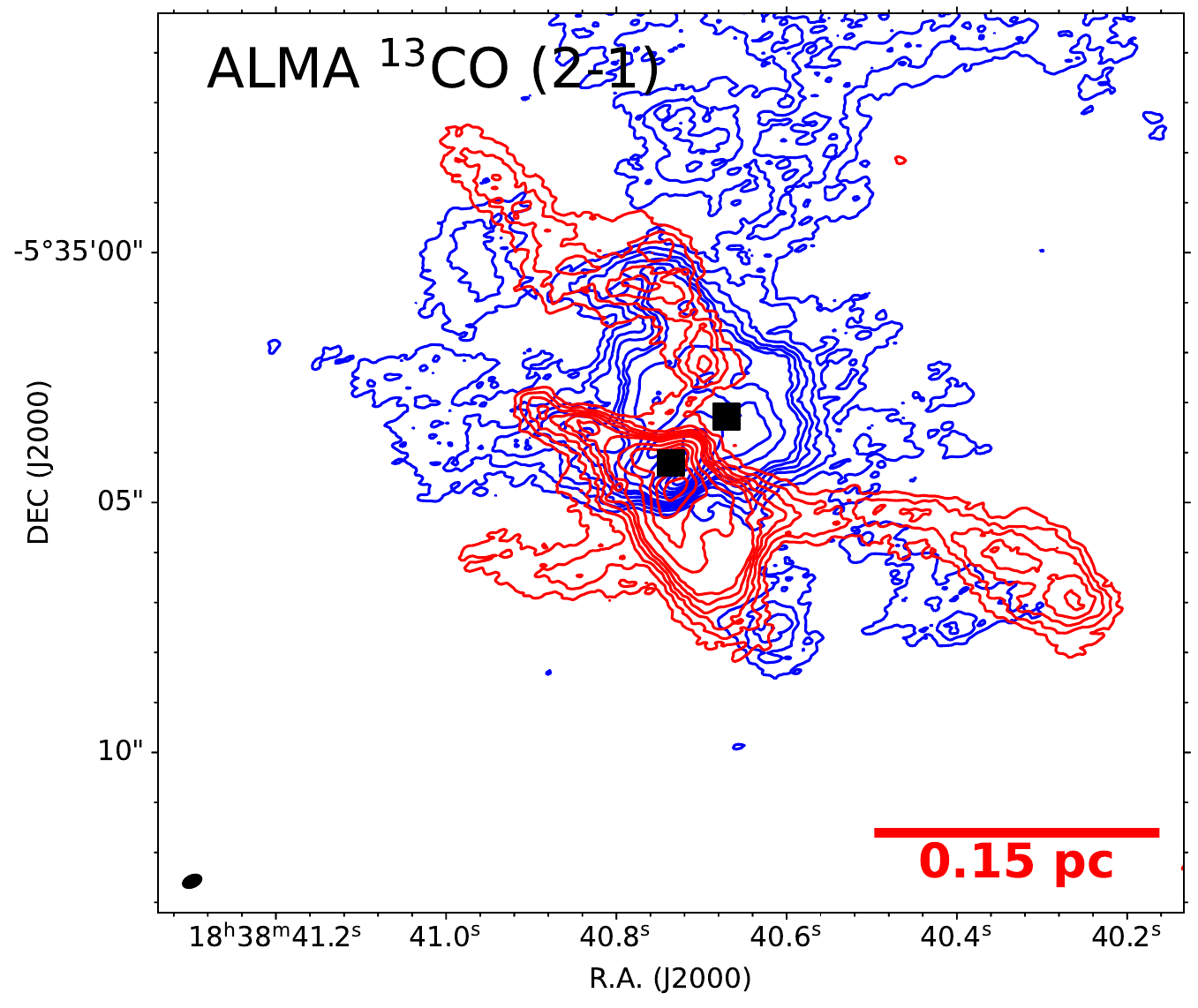}
    \caption{Same as the left panel of Fig. \ref{fig:CO_SiO} but the blue contours are integrated from 97 to 103 $\kms$ and the red ones are integrated from 105 to 111 $\kms$.
    The contour levels of $^{13}$CO (2-1) are set to [8, 12, 16, 20, 24, 28, 40, 52, 64, 76, 88, 100]$\times$ 12.2 m$\Jypb$ $\kms$.}
    \label{fig:COlv}
    \end{figure*}

\end{appendix}

\end{document}